\documentclass[floatfix,secnumarabic,amssymb,nobibnotes,nofootinbib,preprint,aps,pra, superscriptaddress,showpacs]{revtex4-2}
\usepackage{amssymb}
\usepackage[colorlinks=true, citecolor=blue, linkcolor=blue, urlcolor=blue]{hyperref}
\usepackage{times}
\usepackage{amssymb}
\usepackage{amsmath}
\usepackage{cleveref}
\usepackage[utf8]{inputenc}
\usepackage[T1]{fontenc}
\usepackage{mathrsfs}
\usepackage{graphicx}
\usepackage{epsfig}
\usepackage{eucal}
\usepackage{dcolumn}
\usepackage{color}
\usepackage{bm}
\usepackage{graphics,psfrag}
\usepackage{graphicx,psfrag}
\usepackage{braket}
\usepackage{subfigure}
\DeclareUnicodeCharacter{2212}{-}
\DeclareUnicodeCharacter{0393}{+}
\newcommand{\be}{\begin{equation}}
\newcommand{\ee}{\end{equation}}
\newcommand{\bea}{\begin{eqnarray}}
\newcommand{\eea}{\end{eqnarray}}
\newcommand{\ba}[1]{\begin{array}{#1}}
\newcommand{\ea}{\end{array}}

\begin{document}

\title{Manipulating Spin-Lattice Coupling in  Layered Magnetic Topological Insulator Heterostructure \textit{via} Interface Engineering}
\author{Sujan Maity}
\affiliation{School of Physical Sciences, Indian Association for the Cultivation of Science, 2A \&  2B Raja S. C. Mullick Road, Jadavpur, Kolkata - 700032, India}
\author{Dibyendu Dey}
\affiliation{Department of Physics and Astronomy, University of Maine, Orono, ME 04469, USA}

\author{Anudeepa Ghosh}

\affiliation{School of Physical Sciences, Indian Association for the Cultivation of Science, 2A \& 2B Raja S. C. Mullick Road, Jadavpur, Kolkata - 700032, India}

\author{Suvadip Masanta}
\affiliation{Bose Institute, Department of Physics
Main Campus 
93/1, A. P. C. Road
Kolkata - 700 009, India}



\author{Binoy Krishna De}
\affiliation{UGC-DAE Consortium for Scientific Research,
Indore Centre,
University Campus, Khandwa Road, Indore 452001}

\author{Hemant Singh Kunwar}
\affiliation{UGC-DAE Consortium for Scientific Research,
Indore Centre,
University Campus, Khandwa Road, Indore 452001}

\author{Bikash Das}
\affiliation{School of Physical Sciences, Indian Association for the Cultivation of Science, 2A \& 2B Raja S. C. Mullick Road, Jadavpur, Kolkata - 700032, India}
\author{Tanima Kundu}
\affiliation{School of Physical Sciences, Indian Association for the Cultivation of Science, 2A \& 2B Raja S. C. Mullick Road, Jadavpur, Kolkata - 700032, India}

\author{Mainak Palit}
\affiliation{School of Physical Sciences, Indian Association for the Cultivation of Science, 2A \& 2B Raja S. C. Mullick Road, Jadavpur, Kolkata - 700032, India}

\author{Satyabrata Bera}
\affiliation{School of Physical Sciences, Indian Association for the Cultivation of Science, 2A \& 2B Raja S. C. Mullick Road, Jadavpur, Kolkata - 700032, India}

\author{Kapildeb Dolui}

\affiliation{Lomare Technolgies Limited, 6 London Street, London EC3R 7LP, United Kingdom}

\author{Kenji Watanabe}
\affiliation{Research Center for Functional Materials,National Institute for Materials Science, Tsukuba 305-0044, Japan}
\author{Takashi Taniguchi}
\affiliation{International Center for Materials Nanoarchitectonics, National Institute for Materials Science,Tsukuba 305-0044, Japan}
\author{Liping Yu }
\affiliation{Department of Physics and Astronomy, University of Maine, Orono, ME 04469, USA}

\author{A Taraphder}
\affiliation{Department of Physics, Indian Institute of Technology Kharagpur, W.B. 721302, India}

\author{Subhadeep Datta*}
\affiliation{School of Physical Sciences, Indian Association for the Cultivation of Science, 2A \& 2B Raja S. C. Mullick Road, Jadavpur, Kolkata - 700032, India}
\email{sspsdd@iacs.res.in}

\begin{abstract} 
\textbf{Induced magnetic order in a topological insulator (TI) can be realized either by depositing magnetic adatoms on the surface of a TI or engineering the interface with epitaxial thin film or stacked assembly of two-dimensional (2D) van der Waals (vdW) materials. Herein, we report the observation of spin-phonon coupling in the otherwise non-magnetic TI Bi$_\mathrm{2}$Te$_\mathrm{3}$, due to the proximity of FePS$_\mathrm{3}$ (an antiferromagnet (AFM), $T_\mathrm{N}$ $\sim$ 120 K), in a vdW heterostructure framework. Temperature-dependent Raman spectroscopic studies reveal deviation from the usual phonon anharmonicity originated from spin-lattice coupling at the Bi$_{2}$Te$_{3}$/FePS$_{3}$ interface at/below 60 K in the peak position (self-energy) and linewidth (lifetime) of the characteristic phonon modes of Bi$_{2}$Te$_{3}$ (106 cm$^{-1}$ and 138 cm$^{-1}$) in the stacked heterostructure.  The Ginzburg-Landau (GL) formalism, where the respective phonon frequencies of Bi$_{2}$Te$_{3}$ couple to phonons of similar frequencies of FePS$_{3}$ in the AFM phase, has been adopted to understand the origin of the hybrid magneto-elastic modes. At the same time, the reduction of characteristic $T_\mathrm{N}$ of FePS$_3$ from 120 K in isolated flakes to 65 K in the heterostructure, possibly due to the interfacial strain, which leads to smaller Fe-S-Fe bond angles as corroborated by computational studies using density functional theory (DFT). Besides, inserting hexagonal boron nitride within Bi$_{2}$Te$_{3}$/FePS$_{3}$ stacking regains the anharmonicity in Bi$_{2}$Te$_{3}$. Controlling interfacial spin-phonon coupling in stacked heterostructure can have potential application in surface code spin logic devices.  }
\end{abstract}
\maketitle

\section*{Introduction}
The realization of the quantum anomalous Hall effect (QAHE) and topological magneto-electric effect (TMAE) depends on the functional electronic interface between the (anti-)ferromagnet and topological insulator (TI) in the engineered heterostructure framework. Effect on Dirac cone-like surface state of TI due to hybridization and charge transfer across the interface and subsequent formation of an exchange gap due to proximity effects result in an efficient control over the exotic state \citep{PhysRevB.88.144430}. 
Recently, a series of experiments on TIs doped with 3d transition metals \cite{PhysRevLett.108.036805,doi:10.1126/science.1189924, PhysRevLett.109.076801, PhysRevB.103.064428} has lead to the observation of broken time-reversal symmetry (TRS) and the consequent opening of a band gap in the surface band structure. Moreover, magnetically doped-TI shows current-induced giant spin-orbit torque (SOT) which in turn may result in ultra-low power memory and logic devices \cite{9c470cd954594b8696d5bb33a6a47201}.  But still, the sample inhomogeneity and the formation of a disordered cluster in ferromagnetically doped TIs, even with ppm level of doping concentration, restricts the anomalous Hall regime to one or two orders of magnitude less than the Curie-temperature ($T_\mathrm{C}$ $\sim$ 30 K for magnetically doped Bi-Sb alloy) \cite{doi:10.1073/pnas.1424322112}. However, in the interface-controlled magnetic heterostructure (\textit{e.g.} Fe overlayer is deposited on TI), one may preserve long-range ferromagnetism at ambient temperature \cite{doi:10.1021/nl201275q}. Even so, in this metallic FM-TI heterostructure, the possibility of alteration of the topological surface state due to intermixing with the bulk charge carriers, and most importantly, metallic iron film working as a short circuit on TI would impede future spintronic device applications. Furthermore, epitaxial integration (molecular beam epitaxy) of a high-quality thin film of topological insulator (\textit{e.g.} Bi$_2$Te$_3$, Bi$_2$Se$_3$) on insulating magnetic substrate (\textit{e.g.} Y$_3$Fe$_5$O$_{12}$, EuS), with slightest chemical reaction at the interface, can show AHE but limited to low temperature ($\sim$22 K for EuS) due to the choice of the magnet \cite{PhysRevB.88.081407}. Even for antiferromagnets (AFM) exhibiting no macroscopic magnetization, while short-range interfacial exchange coupling \cite{blei2021synthesis} in the interface can be crucial for manipulating magnetic state, as reported in CrSb (AFM with $T_\mathrm{N}$ $\sim$ 700 K)/Cr-doped(Bi, Sb)$_{2}$Te$_{3}$ (FM with $T_\mathrm{C}$ $\sim$ 35 K) thin film superlattice \cite{article}, doping and heterostructure engineering checks the temperature scale close to 50 K. Also, the high sensitivity of the surface states of TI to air exposure puts a limit on the sample characterization and applicability.

For an atomically-flat interface with reasonable air-stability (see Supplementary Information Fig. S19), layered two-dimensional (2D) material-based van der Waals heterostructures (vdWh) are proving to be better to create proximity-induced effects in low-power electronic circuits \cite{Geim_2013}. For example, introducing graphene (Graphene / Europium oxide (EuO)~\cite{PhysRevLett.110.046603}, Graphene/BiFeO$_{3}$ \cite{PhysRevLett.112.116404}, Graphene/EuS\cite{PMID:27019382}, Graphene/yttrium-iron-garnet (YIG) \cite{Leutenantsmeyer_2016}) with magnetic thin films may be used as an electrical read-out of the magnetic order. Moreover, modulation of interfacial spin texture and long-range exchange coupling mediated by magnetic order significantly enhances the transition temperature in the superlattice \citep{PhysRevLett.110.046603}. In fact, the net magnetic moment induced on graphene due to magnetic proximity (0.1 $\mu_{B}$ per C atom) is nearly double that of Pt on magnetic insulator (MI), as reported in the x-ray magnetic circular dichroism (XMCD) studies \citep{Geprags2012}. In the case of vdW semiconductor-magnet heterostructure, or in other words, valley-spin related device consisting of layered transition metal dichalcogenide (TMDC), theoretical predictions have been made where an induced Zeeman field creates giant valley splitting ($\sim$ 300 meV) due to proximity effect in case of (i) ferromagnetic heterostructure in MoTe$_{2}$-EuO \cite{PhysRevB.92.121403}, and (ii) antiferromagnets  as in WS$_{2}$-MnO(111) \citep{Xu:2018PRB}, MoS$_{2}$-CoO(111) \cite{Yang:2018Dec} and TMD-CrI$_{3}$ \citep{PRB:Zollner2019}. As furtherance, the stacked all-2D assembly offers  proximity-induced spin-texture with controlled magnetic anisotropy together with persistent magnetoresistance above transition temperature which has been reported in graphene - vdW FM (Cr$_{2}$Ge$_{2}$Te$_{6}$ with $T_\mathrm{C} \sim$ 60 K) heterostructure \citep{Karpiak:2019}. 

TI-MI heterostructures with vdW materials have been intensively studied using  electronic transport measurements of micro/nano-electronic devices fabricated \textit{via} lithography \cite{lee2016direct}. But, probing the contributions of each constituent layer in the stacking is difficult to conclude from conventional magnetoresistance measurements. Alternatively, Raman spectroscopy offers an indirect tool to detect characteristic phonon mode/s, elementary excitations (\textit{e.g.} magnons), spin-phonon and electron-phonon coupling in 2D magnets (for example, in ferromagnets Fe$_{3}$GeTe$_{2}$ \cite{du2019lattice}, CrSiTe$_{3}$ \cite{milosavljevic2018evidence}, and in antiferromagnets FePS$_{3}$ \cite{ghosh2021spin}) and topological insulators (Bi$_{2}$Te$_{3}$ \cite{buchenau2020temperature}, Bi$_{2}$Se$_{3}$ \cite{irfan2014temperature}) in its few-atomic layer form. In particular, magnetic transition in \textit{strictly-2D} magnets with magnetic anisotropy can easily be detected \textit{via} spin-phonon coupling, as analyzed by Ghosh \textit{et al.}\cite{ghosh2021spin} in case of FePS$_{3}$ by marking the point of deviation (hardening or softening of phonon modes) in temperature-dependent phonon anharmonicity. Note that out of all predicted and experimentally verified 2D FMs and AFMs, FePS$_3$, an antiferromagnet (T$_N$ $\sim$ 120 K in bulk), shows layer-thickness independent transition temperature \cite{ghosh2021spin, vaclavkova2021magnon} which could be crucial for high-temperature applications keeping 2D nature of the heterostructure intact. Also, the pristine quality of less-strained micromanipulated structure with TI which provides low density of states (DOS) at the interface and the greater probability of hybridization with the electronic states from the overlying layered magnets, creates an ideal platform to study the material-specific modification \textit{via} recognizing the Raman modes as a function of temperature. To the best of our knowledge, there are no reports to date that concerns Raman spectroscopic studies on proximity-induced magnetic order on TI in engineered vdW AFM-TI heterostructure. 

Here, we present temperature-dependent Raman spectroscopic studies on \textit{all-2D} topological insulator (Bi$_{2}$Te$_{3}$) - antiferromagnet (FePS$_3$) vdW heterostructure stacked on Si/SiO$_2$ substrate. With temperature (down to 5 K), we track all the characteristic phonon modes of (i) FePS$_3$ which have previously been observed to demonstrate spin-phonon coupling (250 cm$^{-1}$, 280 cm$^{-1}$ and 380 cm$^{-1}$), zone-folding (around 105 cm$^{-1}$), and magnon (120 cm$^{-1}$), and (ii) Bi$_{2}$Te$_{3}$ (106 cm$^{-1}$ and 138 cm$^{-1}$), and analyze the data with the anharmonic phonon-decay model. We note that observed deviation from the usual phonon anharmonicity in peak position (self-energy) and linewidth (lifetime) of the mentioned Raman modes of Bi$_{2}$Te$_{3}$ in the  heterostructure ($T^* \sim$ 60 K) indicates spin-phonon coupling which is, otherwise, not present in an isolated flake. Bi$_{2}$Te$_{3}$ recover its anharmonicity in heterostructure when hexagonal boron nitride (hBN) is placed in between Bi$_{2}$Te$_{3}$ and FePS$_3$.  Also, the characteristic antiferromagnetic  N\'{e}el temperature ($T_\mathrm{N}$) of FePS$_3$ reduces from 120 K (in isolated flake) to 65 K, in the heterostructure possibly due to the interfacial strain which leads to smaller Fe-S-Fe bond angles without varying magnetic anisotropic energy (MAE) which has been investigated using 
density functional theory (DFT) calculations. Spatial control of magnetic proximity-induced spin-phonon coupling in layered stacking of  magnetic topological insulator may open up novel pathways for future gate-tunable all-2D spintronic logic devices.

\section*{Experimental Details}
\label{subsec:exp}
Binary  Bi$_{2}$Te$_{3} $  single crystal was grown through the solid-state reaction method by using box furnace (see the method section for growth details)
\cite{koyano2012single}. Single crystal FePS$_{3} $  was grown by the chemical vapor transport (CVT) method in two zone tube furnace as reported in \cite{https://doi.org/10.48550/arxiv.2208.02729}. Bi$_{2}$Te$_{3}$ crystal was micromechanically exfoliated by standard scotch tape and transferred onto 300 nm Si/SiO$_{2}$ substrate by dry transfer method \cite{castellanos2014deterministic}. Suitable flake was located by using Optical Microscope (OLYMPUS BX53M)) by comparing transparency of the flake.
Heterostructures consist of Bi$_{2}$Te$_{3}$, FePS$_{3}$ and hBN  are fabricated by placing one material onto another by using micromanipulator. The fabricated heterostructures (HS) are: (a) HS-1: Bulk Bi$_{2}$Te$_{3}$ (>500 nm)/ Bulk FePS$_{3}$($\sim$150 nm); (b) HS-2: Bulk Bi$_{2}$Te$_{3}$ ($\sim$ 250 nm)/Bulk FePS$_{3}$($\sim$ 88 nm); (c) HS-3: Bulk Bi$_{2}$Te$_{3}$ ($\sim$140 nm)/ few layer FePS$_{3}$($\sim$10 nm) and (d) HS-4: Bulk Bi$_{2}$Te$_{3}$ (>500 nm) /Bulk hBN(>100 nm)/Bulk FePS$_{3}$ ($\sim$ 150 nm) and (e) HS-5: Bulk Bi$_{2}$Te$_{3}$ (>500 nm) /few layer hBN($\sim$10 nm)/Bulk FePS$_{3}$ ($\sim$ 150 nm) . Room temperature and temperature-dependent Raman measurements of isolated  Bi$_{2}$Te$_{3}$, and these heterostructures were carried out by using Jobin Yvon Horibra LABRAM-HR 800 visible micro Raman system and a 473 nm wavelength radiation from a diode laser as an excitation source. All measurements were performed under high vacuum (10$^{-6}$ mbar) using a liquid helium cryostat (Janis ST-500). Laser beam was focused through a microscope objective with 50x magnification and a spot size of $1\mu$m. Laser power was kept below $ 200 \mu$W  to avoid sample heating.

\section*{Results}
\label{sec:results_discussions}
\begin{figure}[ht]
\centerline{\includegraphics[scale=0.5, clip]{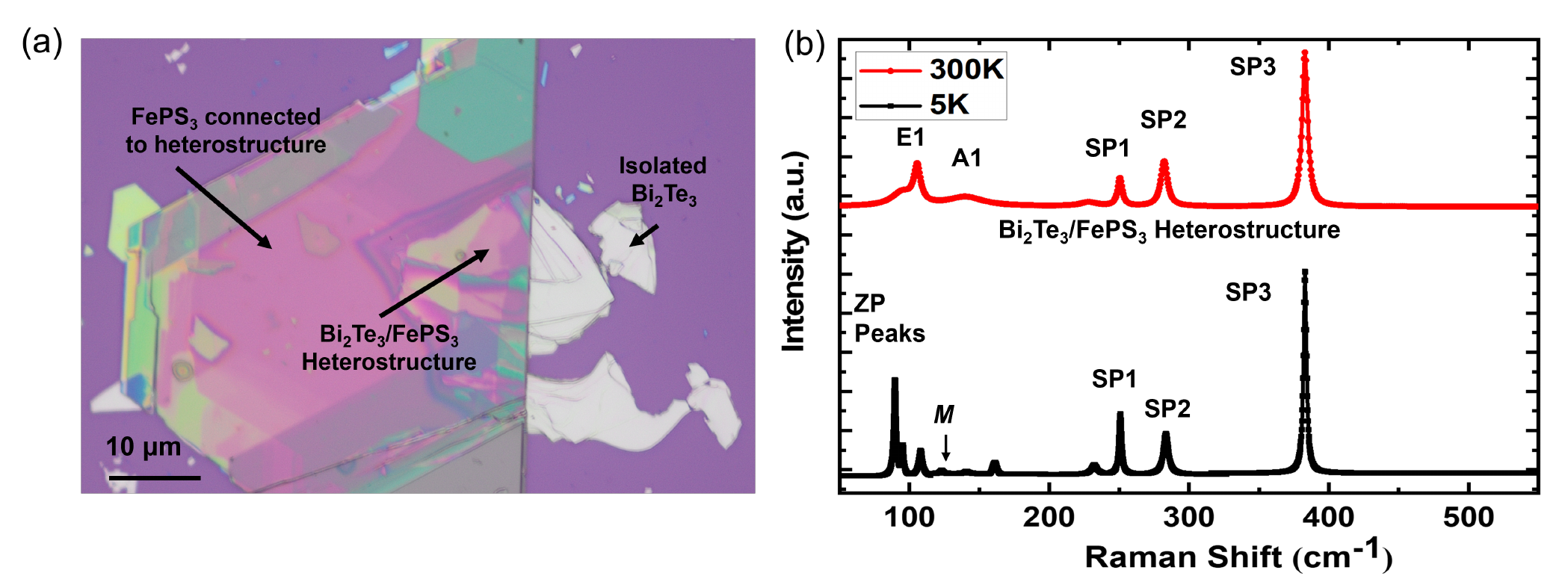}}
\caption{(a) Optical microscopy image of heterostructure (HS-1) fabricated \textit{via} stamping method onto a 300 nm Si/SiO$_{2}$ substrate.  Raman spectroscopy of (b) Bi$_{2}$Te$_{3}$/FePS$_{3}$ heterostructure (HS-1)  at 5 K and room temperature obtained with an excitation wavelength 473 nm. Spin-phonon coupled peaks and  Magnon Peak  are designated as SP(1-3) and $M$ respectively. Zone folded phonon modes (ZPs) appear at low temperature is not present at room temperature. Individual characteristics phonon modes of Bi$_{2}$Te$_{3}$ (denoted as E1 and A1) and FePS$_{3}$ are present in  Bi$_{2}$Te$_{3}$/FePS$_{3}$ heterostructure. \label{fig1}} 
\end{figure}

The crystal structure of Bi$_{2}$Te$_{3}$ contains covalently bonded five mono-atomic planes, which form a quintuple layer (Te-Bi-Te-Bi-Te), weakly bound by van der Waals forces. The dynamical stability of the structure has been confirmed by the absence of imaginary frequencies in the phonon dispersion (see Figure S1).  Among 15 zone center phonon branches (12  optical and 3 acoustic) of  Bi$_{2}$Te$_{3} $, 6 of them are Raman active and 6 are infrared (IR) active modes. The irreducible representation for the zone center phonons can be written as $\Gamma$ = 2E$_{g}$+ 2A$_{1g}$+ 2E$_{u}$+2A$_{1u}$, consistent with the point group symmetry D$_{3d}$ \cite{irfan2014temperature}. Here, in-plane E$_{g}$ and out-of-plane A$_{1g} $ modes are Raman active, whereas E$_{u}$ and A$_{1u} $ are IR active mode \cite{yuan2015raman}.
The phonon dispersion and calculated Raman active modes of the monolayer Bi$_{2}$Te$_{3}$ (see Figure S1b) agree well with observed Raman shifts at 106.19 cm$^{-1}$ (E$_g$) and 138.26 cm$^{-1}$ (A$_{1g}$) from polarization-dependent Raman scattering (see Figure\ref{fig1}(b)).

The irreducible representation of the vibrational modes of the FePS$_3$ monolayer can be written as $\Gamma$ = 8A$_g$+ 6A$_u$+ 7B$_g$ + 9B$_u$ consistent with the C$_{2h}$ point group symmetry of the experimental structure. Among 30 vibrational modes, the A$_g$ and B$_g$ modes are Raman-active  \cite{kargar2020phonon}, three modes are acoustic, and the remaining ones are IR-active. The appearance of negative frequencies (imaginary modes) (see Figure S2a) on the phonon dispersion of the FePS$_3$ monolayer is likely due to the symmetry along the ferromagnetically coupled zigzag chains of Fe atoms \cite{hashemi2017vibrational}. Introducing symmetry-breaking in that direction leads to a strong structural distortion as two Fe-Fe bond lengths along the zigzag chains are no longer equal (2.78 and 3.73 \AA, respectively). This is in contrast to the experimentally determined crystal structures, where ferromagnetically coupled zig-zag chains have a single Fe-Fe bond length of 3.32\AA. Besides, the point group symmetry of this distorted structure is reduced to C$_{s}$, and the lack of centering in the structure doubled the number of vibrational modes of the FePS$_{3}$ monolayer. The phonon dispersion of this low-symmetry distorted structure is shown in Figure S2 (b), and as expected, the imaginary modes no longer exist. The higher wavenumber modes ($\geq$ 250 cm$^{-1}$), depicted in Figure \ref{fig1}(c) as SP1 (250  cm$^{-1} $), SP2 (280  cm$^{-1} $), and SP3 (380  cm$^{-1} $), are mostly attributed to the molecular-like vibrations from (P$_2$S$_6$)$^{4-}$ bipyramid structures, whereas the low-frequency peaks, zone folded phonons ($\leq$ 110 cm$^{-1}$) and magnons ($\sim$ 120 cm$^{-1}$), are from vibrations including Fe atoms \cite{ghosh2021spin, vaclavkova2021magnon}.

Next, we investigated temperature-dependent Raman scattering down to 5 K on HS-1, (see figure \ref{fig1}(a) for reference locations). In case of the heterostructure (HS-1), variation of the classified peaks at low temperature ($T$ = 5 K) from the room temperature data in terms of peak position, full-width-half-maxima (FWHM), and the intensity can be read by the naked eye and may reveal contributions of phonon anharmonicity and spin-phonon coupling (see Figure \ref{fig1}(b)). 

\begin{figure}
\centerline{\includegraphics[scale=0.5, clip]{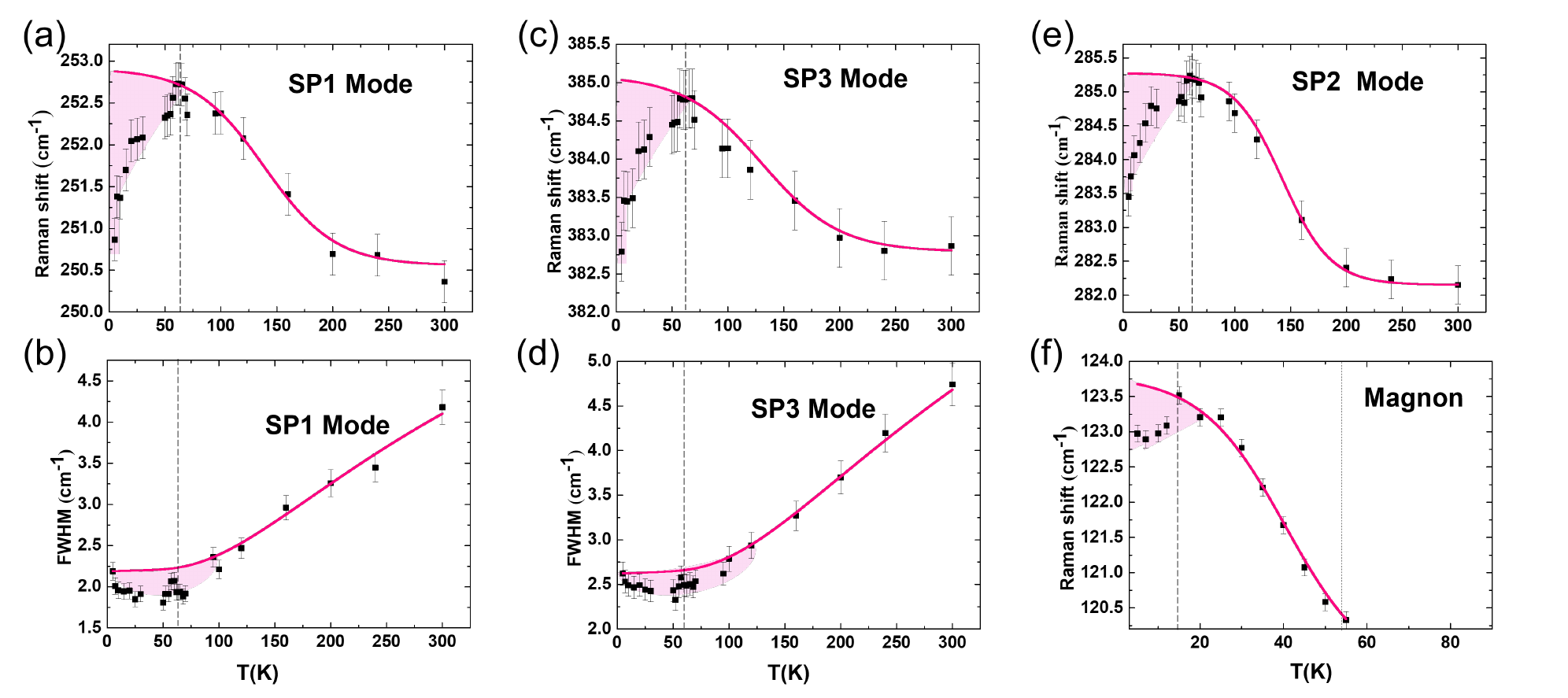}}
\caption{Variation of Peak position and linewidth of FePS$_{3}$ Raman modes  (a), (b) SP1 mode; (c), (d)  SP3 mode; (e) SP2 mode (f) Magnon mode in the heterostructure (HS-1) with temperature. All spin-phonon peaks show phonon anomaly at around 65 K in phonon frequency. At below 65 K, linewidths of all spin-phonon mode of FePS$_{3}$ are not fit with 3-phonon Anharmonic Decay (AD) model due to presence of Bi$_{2}$Te$_{3}$ underneath. One-magnon mode [Fig.(f)] at 120 cm$^{−1}$ at a temperature ($T_{M} \sim$ 60 K) was observed in FePS$_{3}$. Softening of this magnon mode was observed with the temperature and the softening temperature becomes half due to Bi$_{2}$Te$_{3}$ underneath.\label{fig2}}
\end{figure}

\begin{figure}[ht]
\centerline{\includegraphics[scale=0.6, clip]{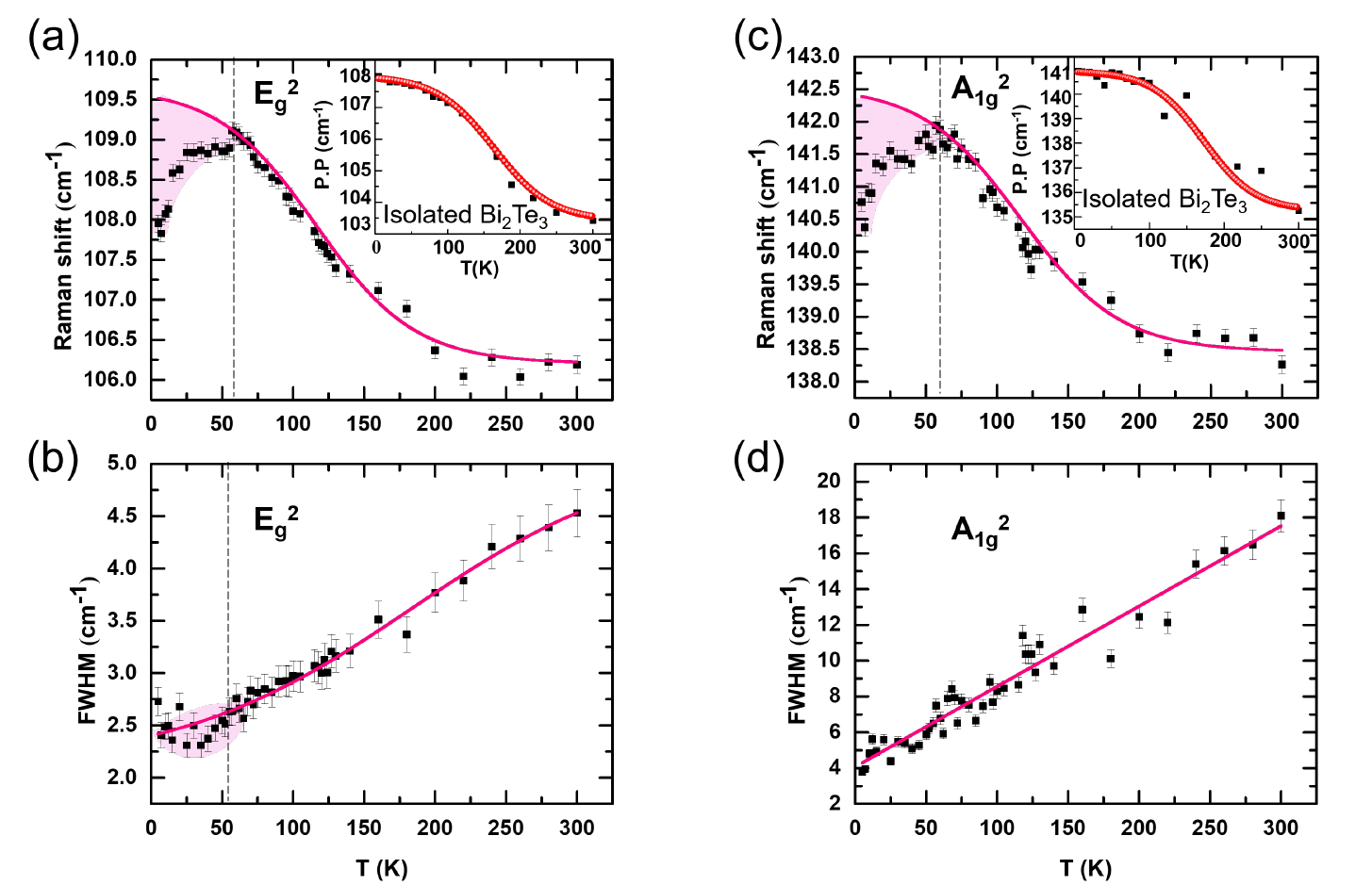}}
\caption{Variation of peak position and linewidth of Raman modes of Bi$_{2}$Te$_{3}$ in the heterostructure (HS-1) (a),(b) In-plane Raman mode  E$_{g}^{2} $ ; (c),(d) Out-of-plane Raman mode  A$_{1g}^{2} $ of Bi$_{2}$Te$_{3} $ with temperature. Insets (a), (c) showing the temperature dependence of Raman modes of isolated  Bi$_{2}$Te$_{3} $. No phonon anomaly was observed in the in-plane and out-of-plane Raman modes of isolated  Bi$_{2}$Te$_{3}$. In case of  Bi$_{2}$Te$_{3}$ in the heterostructure (HS-1), phonon anomaly was observed in both Raman modes of  Bi$_{2}$Te$_{3}$ in phonon frequency at around 60 K. Linewidth of in-plane Raman mode [Fig (b)] of Bi$_{2}$Te$_{3}$ was not fit by Boltzmann sigmoidal anharmonic curve at/below 60 K. Linewidth of out-of-plane Raman mode of  Bi$_{2}$Te$_{3}$ [Fig(d)] increases with the temperature in the whole temperature window. \label{fig3}}
\end{figure}

At the heterostructure region, there may be a significant coupling between the adjacent peaks, as in between broad peak of FePS$_{3}$ (denoted as zone folded phonon (ZP) mode around 108 cm$^{-1}$) and characteristic  E$_{g}^{2}$ mode (designated as E1 in Figure \ref{fig1}(b)) of Bi$_{2}$Te$_{3}$, which are characteristic phonon modes of individual compounds. 
Deconvolution of the hybrid peak around 110 cm$^{-1}$ into two Lorentzian curves confirms that the broad ZP peak of  FePS$_{3} $ becomes sharply peaked (100 times at 5 K) in presence of the underneath Bi$_{2}$Te$_{3}$ layer. Note that the disappearance of the broad peak around 100 cm$^{-1} $ in the overlapping region at low temperature ($\sim$ 100 K) is remindful of the pristine FePS$_{3}$, as reported in \cite{ghosh2021spin}. 
To investigate the temperature dependent phonon behavior, the frequency ($\omega$) and the linewidth
($ \Gamma $) are extracted from the respective Lorentzian fits to specific phonon modes and were fit by symmetric three phonon coupling model (see Figure \ref{fig2}-\ref{fig3}) \cite{menendez1984temperature}: 
\begin{equation}
\omega (T) =\omega_{1}+\frac{\omega^{'}-\omega_{1}}{1+exp[\frac{T-T_{0}}{dT}]}
\end{equation}
\begin{equation}
\Gamma= \Gamma_{0}\left(1+\frac{1}{exp(\frac{hc\bar{\nu_{1}}}{k_{B}T})-1}+\frac{1}{exp(\frac{hc\bar{\nu_{2}}}{k_{B}T})-1}\right)
\end{equation}
\\
Here, the parameters $ \omega^{'} $ and $ \omega_{1} $ represent the top and bottom of the fitted sigmoidal curve, respectively; $ T_{0} $ is the center point, and $dT$ controls the width of the curve. $h$ and $k_\mathrm{B}$ are the Planck and Boltzmann constants, respectively, $c$ is the speed of light and $T$ is the temperature. $\Gamma_{0}$ is the asymptotic value of the linewidth at zero temperature. $\bar{\nu_{1}}$ and $\bar{\nu_{2}}$ are two acoustic phonon modes with different wavenumbers with opposite wavevector.

In the isolated  Bi$_{2}$Te$_{3}$ (see the inset of Figure \ref {fig3}(a)), no phonon anomaly was observed in the peak position and linewidth of the phonon modes, possibly due to the dominant phonon-phonon scattering \cite{buchenau2020temperature}. The observed anharmonic behaviour is independent of  the excitation wavelength \cite{buchenau2020temperature}, different structural form (bulk single crystal \cite{mal2019vibrational}, nanowires, nanoribbons \cite{park2016thermal} etc.), and the thickness \cite{buchenau2020temperature, singh2021study} of  Bi$_{2}$Te$_{3}$ (see Supplemental Table VIII). The behavior can simply be described by the above mentioned symmetric three-phonon coupling model \cite{klemens1966anharmonic} where a zone centre optical phonon decays into two acoustic phonons with equal energies and opposite momenta. Note that the fitting of the temperature dependent peak positions of the in-plane and out-of- plane phonon modes of isolated Bi$_{2}$Te$_{3}$ (see inset of Figure \ref{fig3}(a),(c)) and  FePS$_{3} $ connected to the heterostructure (HS-1) is depicted in the Supplementary Information (Figure S3) \cite{casto2015strong}. 



In case of FePS$_{3}$, in the heterostructure (HS-1), at/around 65 K, SP1-SP3 modes show deviation from the anharmonic fit for each of the phonon frequencies [see Figure \ref{fig2}(a),(c),(e)]. Note that antiferromagnetic ordering sets in ($T_\mathrm{N}$) at 120 K, irrespective of the thickness of the FePS$_{3}$ flake \cite{ghosh2021spin}. Also, no such changes were detected for the linewidth, may be due to the insignificant contribution of spin-phonon interaction to the phonon relaxation process \cite{prosnikov2018magnetic}. The nature of the temperature dependent linewidth also suggests that the  effect of phonon-impurity scattering, and electron-phonon interaction can be ruled out, as discussed in \cite{irfan2014temperature}, whereas the violation of three phonon anharmonic decay model was observed in the linewidth of FePS$_{3}$ spin-phonon coupled modes (see figure \ref{fig2}(b),(d)) in the heterostructure due to Bi$_{2}$Te$_{3}$ underneath. 
\\
\\  

Noticeably, in the Bi$_{2}$Te$_{3}$/FePS$_{3}$ heterostructure (HS-1), the enhancement of the intensity  in all the Raman modes of FePS$_3$, compared to the isolated flake, can be explained by the electron transfer at the Bi$_{2}$Te$_{3}$/FePS$_3$ interface. 
Adopting a simple model where the interface has been treated as a metal (Bi$_{2}$Te$_{3}$ is sufficiently n doped with $ n \sim  10^{19}$ cm$^{-3}$) \cite{okuda2001large} - semiconductor (FePS$_3$ with bandgap 1.60 eV) junction, band alignment, depicted in Figure S4, indicates that the electrons will be transferred from FePS$_3$ into the Bi$_{2}$Te$_{3}$. The barrier ($E_{bar} \sim$ 1.17 eV) formed at the interface, which is the difference between the work functions of the Bi$_{2}$Te$_{3}$ ($\phi_{BT}$ $\sim$ 5.30 eV) \cite{lee2016enhanced} and the FePS$_3$ ($\phi _{FPS}$ $\sim$ 4.13 eV), is less than the excitation wavelength (2.62 eV). Consequently, available transitions for the Raman scattering may involve real energy levels which drastically enhances the intensity of the characteristic Raman modes of FePS$_3$ \cite{PhysRevB.98.014308}. 


One-magnon mode ($M$) at 120 cm$^{-1}$ at a temperature ($T_{M}\sim$ 60 K) was observed in FePS$_{3}$. Softening of this magnon mode appears at 30 K with $\bigtriangleup \omega \sim$ 4 cm$^{-1}$ in FePS$_{3}$ connected to heterostructure (HS-1) (see figure S5(a)) while, this anomaly occurs at 15 K [Fig.\ref{fig2}(f)] with very small $\bigtriangleup\omega \sim $ 0.56 cm$^{-1}$ due to Bi$_{2}$Te$_{3}$ underneath, in the heterostructure \cite{ghosh2021spin, vaclavkova2021magnon}. Additionaly, magnetic field dependent Raman scattering at 4 K offers an insightful observation on the spliiting of the magnon mode in the heterostructures. By applying a magnetic field perpendicular to the $ab$ plane of the HS-1,  magnon-gap excitation can be splitted into two components (linear in B with $g$-factor close to 2.15) following the easy-axis antiferromagnetic properties which has already been  reported  by  Vaclavkova \textit{et al.} for the isolated FePS$_{3}$ \cite{vaclavkova2021magnon}.  Moreover, no such changes in the strength of magnon-phonon coupling ($\sim$ 3 cm$^{-1}$) was observed in the case of heterostructure. In due course, while applying magnetic field parallel to  the $ab$ plane of the HS-1, magnon mode of FePS$_{3}$ splits at much lower magnetic field  ($\sim$ 9 T) compared to in-plane splitting field in isolated FePS$_{3}$ ($\sim$ 16 T) reported in \cite{vaclavkova2021magnon}. Note that the Raman modes of isolated Bi$_{2}$Te$_{3}$ shows no magnetic field dependence in either configuration, as also reported in previous studies \cite{buchenau2020temperature} (see Supplementary Information Figure S24, Table IX) . To understand the reduction of the external {\textquotedblleft in-plane\textquotedblright} magnetic field required for the splitting of the magnon in the HS-1, one can speculate the effect of the time-independent antidamping torque  which may generate further {\textquotedblleft in-plane\textquotedblright} field-like effect predicted for antiferromagnetic topological insulators with preserving the gapless states in TI \cite{PRBMTI} (see Supplementary Information Figure S24 , Table IX). Besides, to inspect the origin of the possible magnon temperature reduction in hybrid FePS$_\mathrm{3}$, the spin wave stiffness constant ($D$) which is linked to the long wavelength limit of the acoustic mode of magnon dispersion can be determined. Our noncollinear density functional theory (ncDFT) calculations (see the methodology for computational details) predict that monolayer FePS$_\mathrm{3}$ has spin wave stiffness of 245 meV\AA$^2$. Interestingly, $D$ value gets reduced to 158 meV\AA$^2$ for FePS$_\mathrm{3}$ deposited onto six quintuple layer (6QL) of Bi$_\mathrm{2}$Te$_\mathrm{3}$.

Figure \ref{fig3} presents the temperature dependence of the in-plane and the out-of-plane phonon modes of Bi$_{2}$Te$_{3}$ in the heterostructure. Unlike pristine Bi$_{2}$Te$_{3}$ (see the inset of Figure \ref{fig3}a and \ref{fig3}c) \cite{buchenau2020temperature}, at/around 60 K, respective phonon modes,  E$_{g}^{2} $ (see Figure \ref{fig3}a) and A$_{1g}^{2} $ (see Figure \ref{fig3}c), show clear deviation ($\bigtriangleup\omega_{5 K}$) of 1.56 cm$^{-1}$ and 1.62 cm$^{-1}$ from the anharmonic fit, respectively. Regarding linewidth, while E$_{g}^{2}$ mode shows slight deviation ($\sim$  0.56 cm$^{-1}$) in the phonon behaviour around 60 K (see Figure \ref{fig3}b), no such abrupt changes were observed for  A$_{1g}^{2}$ mode (see Figure \ref{fig3}d). Theoretically, spin-phonon coupling can be introduced as spin susceptibility ($\chi_M$) in the phonon self energy (SE) which consists of the real (frequency shift) and the imaginary part (line broadening)  ($\Delta(\omega_{j}(q),T)+i\Gamma(\omega_{j}(q),T)$), analogous to modification due to electron-phonon coupling \cite{buchenau2020temperature}. On account of this, from our result, a clear departure of $\chi_M$ from the usual phonon behavior in the Bi$_{2}$Te$_{3}$ can be identified at/around 60 K in the heterostructure (see Figure S6(a), (b)). 
Thickness dependent Raman measurements were done on Bi$_{2}$Te$_{3}$/FePS$_{3}$ heterostructures (HS-2, HS-3) with temperature ranging from 5 K to 300 K. The coupling strength was observed to be high for the heterostructure consisting of thick-layer Bi$_{2}$Te$_{3}$ and FePS$_{3}$. Strength of spin-phonon coupling of Raman modes of Bi$_{2}$Te$_{3}$ decreases with the reduction of thickness (see fig. S8) of individual material as observed in \cite{PRMTI-AFM}. In the heterostructure, with reduced thickness, surface phonon polariton is red shifted and as a result, it becomes off-resonant with antiferromagnetic magnon, which, in-turn, reduces the coupling strength of the hybridized quasi-particle. The deviation from the phonon anharmonicity $\Delta\omega$ (related to square of magnetization) decreases with lowering the thickness of Bi$_{2}$Te$_{3}$ and FePS$_{3}$ in heterostructures. In the case of HS-1, the value of $\Delta\omega$ is close to 1.21 which is higher than   the recorded values for HS-2 ($\Delta\omega$ $\sim$ 0.69) and HS-3 (($\Delta\omega$ $\sim$ 0.57) for the in-plane and out-of-plane Raman modes of Bi$_{2}$Te$_{3}$. The deviation ($\Delta\omega$) is more prominent in out-of-plane mode (A$_{1g}^{2}$) of Bi$_{2}$Te$_{3}$ in case of few layer heterostructure (see fig. S8(d)) since Bi and Te atoms vibrate perpendicularly to the layer surface for A$_{1g}^{2}$ mode \cite{Nano-Bi2Te3}. The coupling strength decreases but the characteristic temperature for spin-phonon coupling of Bi$_{2}$Te$_{3}$ remains invariant ($\sim$ 60 K) with reduction of thickness (see fig. S8). The robustness of spin-phonon coupling of Bi$_{2}$Te$_{3}$ in Bi$_{2}$Te$_{3}$/FePS$_{3}$  heterostructure (HS-4, HS-5) could be destroyed by inserting hexagonal boron nitride (hBN) in between Bi$_{2}$Te$_{3}$ and FePS$_{3}$  (see Figure \ref{fig5}). The insulating nature of the FePS$_{3}$ with highly transparent hBN \cite{nguyen2020visibility} in the middle of the  stacked assembly make the laser penetration depth very large (also recognizable under optical microscope), which in turn facilitates the recording of the characteristic Raman modes of individual materials even in bulk thicknesses (see Figure \ref{fig5}(b)). No such phonon anomaly (independent of the thickness of hBN) was observed in the in-plane and out-of-plane Raman modes of Bi$_{2}$Te$_{3}$ (see Figure \ref{fig5}(c), (d)), similar to the case of isolated Bi$_{2}$Te$_{3}$.

 Typical example of magnetic proximity effect \cite{tang2020magnetic, ghiasi2021electrical, PhysRevB.108.L100406, chen2022control} is the extension of the spin order beyond the interface of diluted magnetic semiconductors (DMS), namely (Ga, Mn)As, and Fe overlayer through an anti-parallel alignment. Avoiding the effect of diffusion/seggregation in DMS subsystem, nontrivial surface state of TI in contact with FMI can be understood from the propagation of spin-polarized charge carriers resulting in building up finite spin polarization close to the interface \cite{maccherozzi2008evidence}. Moreover, presence of the TI's conducting surface states is crucial for the spin polarized charge transfer \cite{liu2015enhancing, chen2022control}. On a related note, diffusion of the magnetic atoms (\textit{e.g} Eu atoms in EuS) into the first QL of TI (\textit{e.g} Bi$_2$Se$_3$) during the growth of the heterostructure can be the origin of “interface ferromagnetism”, as reported in  \cite{eremeev2015interface}. But, in the concerned study, heterostructure has been formed \textit{via} stamping of one vdW material onto another, eliminating the possibility of diffusion which usually happens during the thin film growth \cite{liu2017wafer}. Interfacial ferromagnetism remains evident up to room temperature in Bi$_{2}$Se$_{3}$/EuS heterostructure. Significant enhancement of T$_{C}$ was observed at the interface arising from large spin–orbit interaction and spin–momentum locking of the topological insulator surface \cite{katmis2016high}.  In Bi$_{2}$Te$_{3}$/Fe$_{3}$GeTe$_{2}$ heterostructure, T$_{C}$ enhances up to 400 K, which is attributed to the interfacial exchange coupling effect between Bi$_{2}$Te$_{3}$ and Fe$_{3}$GeTe$_{2}$, which induces the enhancement of T$_{C}$ for the 2D ferromagnetism \cite{wang2020above}. The enhancement of T$_{C}$ of Cr$_{2}$Ge$_{2}$Te$_{6}$ from 65 K to 150 K was observed in W/Cr$_{2}$Ge$_{2}$Te$_{6}$ heterostructure  due to the formation of W-Te bonding at the interface \cite{zhu2022interface}.
In the stacked structure, strain engineering may enhance the $T_\mathrm{C}$ by raising the exchange energy between the cation sites which may be mediated by the polarized anion atoms \cite{wang2020above}. But even for the fully epitaxial vdW heterostructure (\textit{e.g} Fe$_x$Cu$_{1−x}$Se and Bi$_2$Te$_3$), lattice mismatch, as large as 20\%, may occur through van der Waals bonding across the interface \cite{ghasemi2017van}. For a typical sample size of sub-micrometer length scale, structural, chemical and electronic analysis of the heterointerface as formed in the vdW stacking, can be probed by low-energy electron microscopy (LEEM), which is beyond the scope of the current work \cite{PhysRevB.89.155408, das2021manipulating}. However, in our micromanipulated vdW Bi$_{2}$Te$_{3}$/FePS$_3$ heterostructure, possible strain has been calculated as low as 0.5\%. 

\begin{figure}[h]
\centerline{\includegraphics[scale= 0.6, clip]{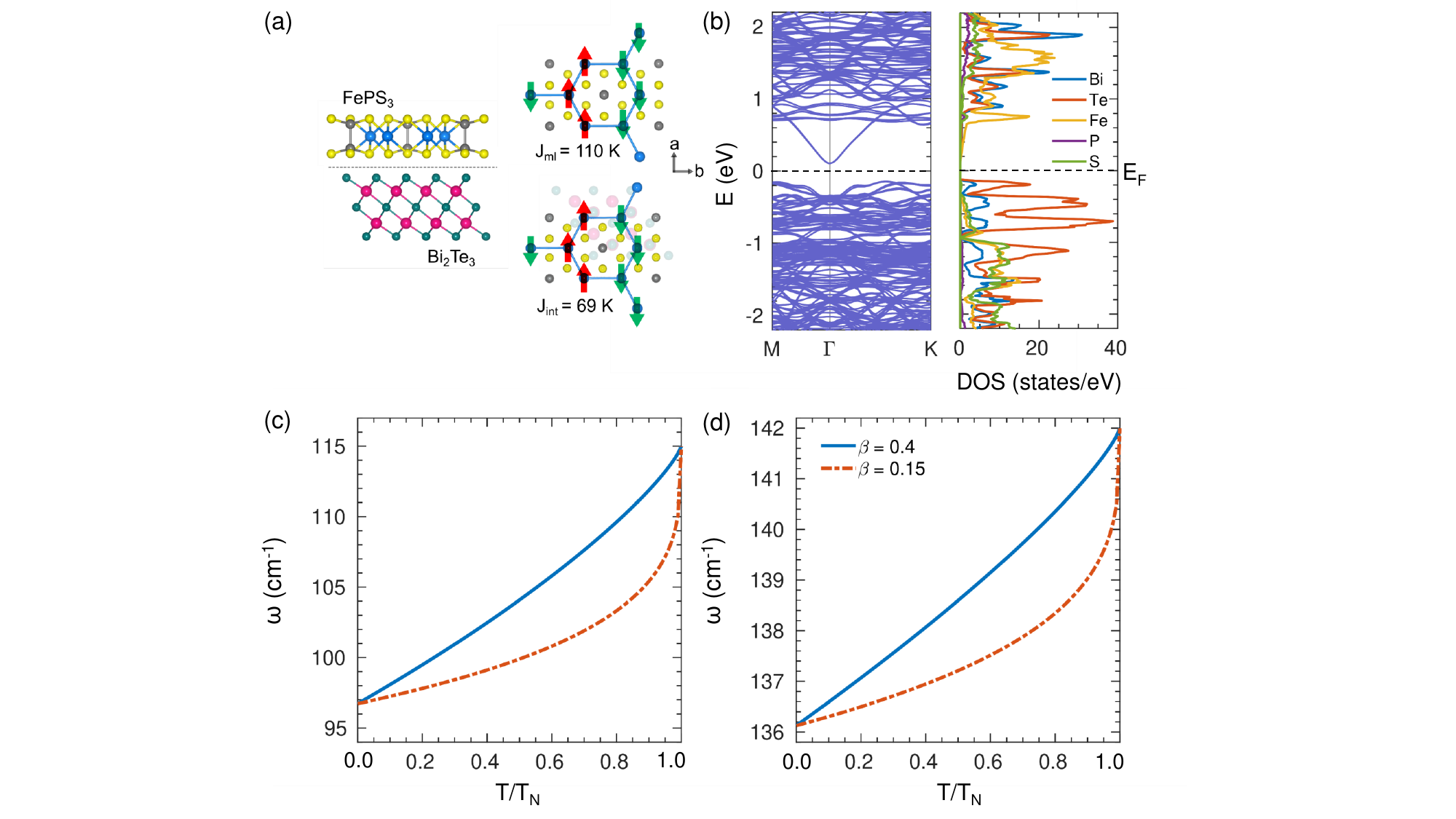}}
\caption{(a) A schematic of Bi$_{2}$Te$_{3}$/FePS$_{3}$ heterostructure and the zigzag AFM order of an isolated FePS$_{3}$ monolayer and a monolayer on top of Bi$_{2}$Te$_{3}$. The respective AFM exchange values are also shown. (b) Band structure and atom projected density of states (DOS) of Bi$_{2}$Te$_{3}$/FePS$_{3}$ heterostructure obtained within GGA+U+SOC reveal that the system is a narrow band gap semiconductor. Temperature dependence of (c) 115 cm$^{-1}$ and (d) 141 cm$^{-1}$ FePS$_{3}$ phonon modes that are coupled to Bi$_{2}$Te$_{3}$ modes are shown for two critical exponent ($\beta$) values obtained from experiments. \label{fig4}}
\end{figure}

\begin{figure}[h]
\centerline{\includegraphics[scale=0.6, clip]{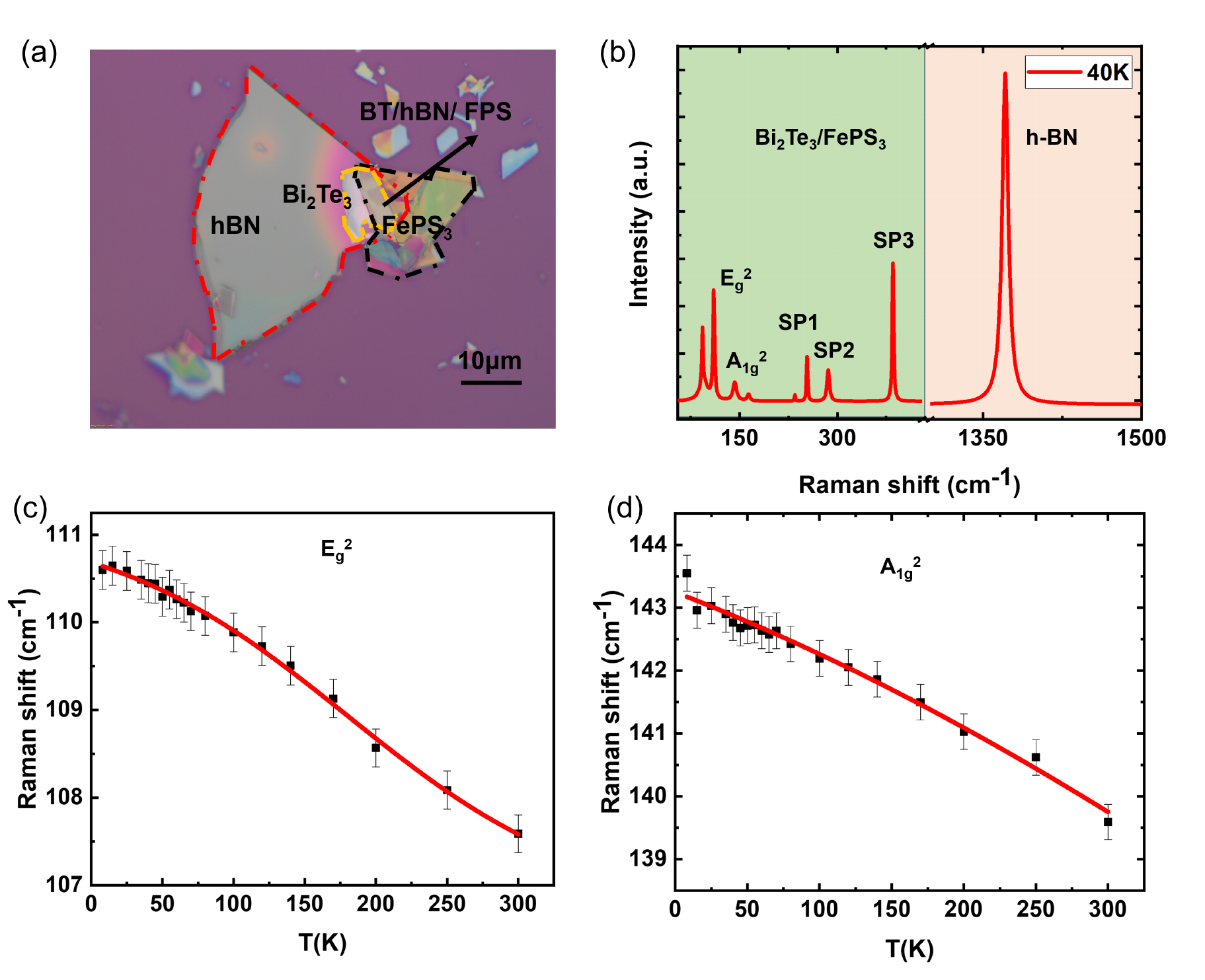}}
\caption{(a) Optical microscopy image of  heterostructure (HS-4). Orange dotted line surrounds Bi$_{2}$Te$_{3}$ flake. Bulk hBN (red dotted line) is first placed on top of Bi$_{2}$Te$_{3}$. FePS$_{3}$ (black dotted line) covers both Bi$_{2}$Te$_{3}$ (some portion)  and hBN (some portion)  resulting the formation of Bi$_{2}$Te$_{3}$/hBN/ FePS$_{3}$ heterostructure. (b) Raman spectra of the heterostructure (HS-4) at low temperature. Individual Raman modes of Bi$_{2}$Te$_{3}$, hBN and FePS$_{3}$  are appeared in Bi$_{2}$Te$_{3}$/hBN/ FePS$_{3}$ heterostructure .   Variation of peak position of Raman modes of Bi$_{2}$Te$_{3}$   (c) In-plane Raman mode  E$_{g}^{2} $ ; (d) Out-of-plane Raman mode  A$_{1g}^{2} $ of Bi$_{2}$Te$_{3} $ with temperature in the heterostructure (figure (a)). No phonon anomaly was observed in the in-plane and out-of-plane Raman modes of Bi$_{2}$Te$_{3}$ due to the presence of hBN like isolated one. \label{fig5}}
\end{figure}

The magnetic and electronic properties of the Bi$_{2}$Te$_{3}$/FePS$_{3}$ heterostructure have been investigated by performing DFT calculations using a supercell containing both individual layers coupled weakly by the vdW interaction (see methods for details). Since the point group symmetry (D$_{3d}$) remains invariant with the thickness of Bi$_{2}$Te$_{3}$, we have considered Bi$_{2}$Te$_{3}$ monolayer for our calculations to save computational time. Note that no significant Raman shift was observed in the in-plane phonon characteristic mode (E$_{g}^{2}$) of Bi$_{2}$Te$_{3}$, whereas, out-of-plane Raman mode (A$_{1g}^{2}$) shows a maximum shift of 6 cm$^{-1}$ with thickness \cite{Nano-Bi2Te3, shahil2012micro,shahil2010crystal} (See Supplementary Information Table IV, Table VII). In the context of the current study, temperature dependent phonon anomaly as detected from the deviation from anharmonicity, in phonon frequency for a certain thickness is conceptually different from the thickness dependent variation of Raman shift. Contrastingly, all phonon modes of FePS$_{3}$ hardly show any Raman shift with decreasing thickness down to monolayer (see Supplementary Information, Figure S23, Table VI). It is worth mentioning that FePS$_{3}$ exhibits layer thickness independent transition temperature (T$_{N} \sim $ 120 K) from bulk to monolayer \cite{lee2016ising}. In this framework, the S-termination of the FePS$_{3}$ layer (on top of the Bi$_{2}$Te$_{3}$ layer) is energetically the most stable configuration, and the stoichiometry of each individual layer is also maintained. Interestingly, the magnetic ground state (z-AFM order; cf. (Fig. \ref{fig4}a) of the FePS$_{3}$ layer remains unchanged even in this heterostructure set-up. The robustness in AFM order of FePS$_{3}$ compound \cite{lee2016ising} from bulk to monolayer to heterostructure is promising for device applications. We obtain $\sim$ 3.5 $\mu_{B}$ magnetic moment at the Fe$^{2+}$ sites, whereas the other atoms have negligibly small moments (< 0.1 $\mu_{B}$). Furthermore, the nature of the spin anisotropy of the interface FePS$_{3}$ layer was determined from our calculated magnetocrystalline anisotropy energies (MAE). The obtained MAE (E$_{in-plane}$-E$_{z}$) value 1.31 meV/f.u. of the FePS$_{3}$ layer indicates the easy axis (out-of-plane) of  magnetic anisotropy of the system. Relatively large MAE value \cite{PRMDD} originates from the strong spin-orbit coupling (SOC) in the Bi$_{2}$Te$_{3}$/FePS$_{3}$ heterostructure. In contrast, magnetic shape anisotropy (MSA), which represents the anisotropic dipolar interaction of free magnetic poles and tends to align magnetic moments parallel to surfaces, only plays a significant role when SOC is weak \cite{PRMDD}. MSA values are generally in the order of µeV, and thus, they are negligible for Bi$_{2}$Te$_{3}$/FePS$_{3}$ heterostructure . Therefore, the long-range order observed in the Bi$_{2}$Te$_{3}$/FePS$_{3}$ heterostructure is primarily governed by MAE. The GGA+$U$+SOC electronic structures (Fig {\ref{fig4}b}) of the Bi$_{2}$Te$_{3}$/FePS$_{3}$ heterostructure reveal that the system is a narrow band-gap (0.25 eV) semiconductor. 
The atom-projected density of states (DOS) shows that the bands below the Fermi level (FL) are predominantly occupied by Bi and Te valence states, and there is a strong hybridization between the Fe $d$ levels and the valence-band orbitals of Bi$_2$Te$_3$. Since this hybridization is
spin-dependent, the proximity exchange could be significant~\cite{prox-theory} and may lead to gaped surface states as opposed to the metallic surface states of isolated Bi$_{2}$Te$_{3}$ (see Figure  S5(b). Note that scanning tunnelling spectroscopy on Bi$_{2}$Te$_{3}$/FePS$_{3}$ heterostructure may invoke insights into the atomistic origin of the proximitized magnetic ordering which is beyond the scope of the current work.



To understand the microscopic origin of the reduction of $T_\mathrm{N}$ of FePS$_{3}$ in the heterostructure, we calculate the interface AFM exchange ($J_{int}$) by using the energy difference between FM and z-AFM spin configurations after mapping them to the Heisenberg model $H=J_{int}$ $\displaystyle\sum _{i>j}S_{i}\cdot S_{j}$ (cf. supplemental materials S7 for details). We estimate $J_{int}$ to be 69 K, lower than the AFM exchange $J_{ML}$ of the FePS$_{3}$ monolayer (110 K). The lowering of AFM exchange in the Bi$_{2}$Te$_{3}$/FePS$_{3}$ heterostructure is consistent with the experimentally observed trend in $T_\mathrm{N}$. In FePS$_{3}$, Fe atoms are in an edge-sharing octahedral environment with their neighboring S, and in this case, the Fe-S-Fe bond angle is close to 90$^o$. In this situation, direct d-d hopping, which gives rise to an antiferromagnetic exchange competes with the FM superexchange \cite{goodenough, kanamori}. In the 0.5\% bi-axial strained structure of the FePS$_{3}$ monolayer, the Fe-S-Fe bond angle is $\sim$10$^o$ smaller than the unstrained structure, whereas the Fe-Fe bond length remains almost the same. Due to this reason, the effective AFM interaction ($J_{ML}$ = 80 K) of the strained FePS$_{3}$ monolayer gets weakened by the enhancement of FM superexchange. In Bi$_{2}$Te$_{3}$/FePS$_{3}$ monolayer, we also observe a similar trend with higher reduction in Fe-S-Fe bond angles , and thus the overall AFM exchange becomes much weaker. Therefore, the interfacial strain which leads to smaller Fe-S-Fe bond angles could primarily attribute to the lowering of $J_{int}$ in the heterostructure system. The nature of the strain demonstrated in the calculations corresponds to compressive biaxial strain for the FePS$_{3}$ monolayer. Similar strain effects due to the lattice mismatch at the vdW interface between Bi$_{2}$Te$_{3}$ and FePS$_{3}$ was observed (See Supplementary Information Table V). In case of FePS$_{3}$ connected to the heterostructure, reduction of T$_{N}$ was observed at/around 65 K (see supplementary information Fig. S3). The interfacial strain arises due to the lattice mismatch, which propagates laterally through the layer at a certain distance ($\sim$ 100 nm) from the edge of the interface and decreases with the distance as reported in previous computational studies \cite{smolyanitsky2011simulation}. Similar hetero-bonding effect due to strain has been detected in  WSe$_{2}$-MoS$_{2}$  heterostructure \textit{via} scanning probe microscopy \cite{zhang2018strain}. In our case, the strain propagation is the most probable origin of the reduced T$_{N}$ in FePS$_{3}$ connected to the heterostructure (beyond overlapping region), as depicted in the supplemental figure (Fig. S3). A proper mapping of the two-dimensional strain tensor (in-plane) would be necessary to understand the precise extension of the decay. Further studies, like spin-polarized scanning probe microscopy, are required for an atomistic picture of strain distribution. In the current work, in spite of the reduction in T$_N$, increment of the sublattice magnetization of FePS$_{3}$ in case of the heterostructure can be observed as $\Delta\omega$ value is enhanced by 4 times compared to isolated FePS$_{3}$ which may result in better device performance, may be realized \textit{via} magneto-transport measurement.

A sharp decrease in the Raman shift is observed at 30 K (see figure \ref{fig3}(a),(c)) in  both modes of Bi$_{2}$Te$_{3}$ in the heterostructure. This could be understood from the temperature-dependent antiferromagnetic order parameter. The magnetization of the bulk state falls faster than the surface states with temperature (see figure S9 (a),(b)) as reported in \cite{PhysRevB.85.195119}. In mean field approximation, one can relate $\Delta\omega$ to magnetization as $\Delta\omega(T)\propto \frac{M^{2}(T)}{M_{max}^{2}}$. Magnetization ($M$) is plotted with temperature for the phonon modes of Bi$_{2}$Te$_{3}$ in the heterostructure (see figure S9 (a),(b)) and fit with $M\approx |(1-\frac{T}{T_{N}})|^{\beta} $ equation, where $\beta$ is the critical exponent. $\beta$ value at 60 K (0.15) corrborates with 2D Ising model, the origin of the surface magnetic contribution comes from FePS$_{3}$ and the $\beta$ value at 30 K (0.35) corresponds to 3D Heisenberg model, responsible for bulk contribution. The exponent value, $\beta$ for isolated FePS$_{3}$ and FePS$_{3}$ in the heterostructure are close to the mean field (see figure S9(c)). With decreasing the Bi$_{2}$Te$_{3}$ layers, proximity induced bulk magnetization (related to $\Delta\omega$) in Bi$_{2}$Te$_{3}$ decreases and as a result coupling strength at the interface was also reduced. Induced bulk magnetization reduces more than 4 times in Bi$_{2}$Te$_{3}$ when thickness of Bi$_{2}$Te$_{3}$ reduces from $\sim $ 500 nm (HS-1) to $\sim$ 12 nm (HS-II) (see supplementary figure, Fig. S25). 

Now we turn to discuss the spin-phonon coupling of the Bi$_{2}$Te$_{3}$/FePS$_{3}$ heterostructure. This complex heterostructure has a P$_1$ space group and C$_1$ point group symmetry, and due to having a large number of atoms in the supercell, the phonon calculation of this heterostructure is very expensive. Therefore, we calculate the phonons of individual monolayers and combine them together to understand the Raman data of the Bi$_{2}$Te$_{3}$/FePS$_{3}$ heterostructure. The spin-phonon coupling parameter of the FePS$_{3}$ monolayer has been estimated from the shift in $\Gamma$-phonons due to a change in magnetic order from z-AFM to FM. The Raman active modes corresponding to both magnetic order and relative change in phonon frequencies ($\Delta^{rel}_\lambda = \frac{\omega_{FM}-\omega_{z-AFM}}{\omega_{z-AFM}}$) are listed in Supplemental Table 2. The significant change in phonon frequencies due to a change in magnetic order indicates the strong coupling between magnetic and lattice degrees of freedom in the FePS$_{3}$ monolayer.

In particular, 115 cm$^{-1}$ and 141 cm$^{-1}$ A$_{g}$ modes of FePS$_{3}$ show significant spin-phonon couplings. Interestingly, the frequencies of these two modes are close to the phonon frequency of E$_g$ and A$_{1g}$ modes of Bi$_{2}$Te$_{3}$ (Supplemental Table 1). There is a possibility that the Bi$_{2}$Te$_{3}$ modes would couple to phonons of similar frequencies of FePS$_{3}$
due to resonance, and the hybrid modes appear in the Raman spectrum. Such coupling phenomena were reported in earlier literature \cite{dey2020phonon}, where magnon excitations with Raman-allowed symmetries couple to similar frequency phonon modes showing strong spin-phonon coupling. In order to illustrate the temperature dependence of the hybrid modes, we consider the Ginzburg-Landau (GL) theory formalism \cite{dey2020phonon}. Due to the spin-phonon coupling, the phonon frequency will vary with temperature below $T_\mathrm{N}$ as the z-AFM order sets in, and the modified phonon frequency is given by the formula:
\begin{equation}
\omega_{\lambda}= 2\bigtriangleup_{\lambda}m^{2}+ \sqrt{\omega_{0\lambda}^2+4\bigtriangleup_{\lambda}^2m^4}
\end{equation}
where $m$ $\sim (1-\frac{T}{T_{N}})^\beta$ is the magnetic order parameter, $\omega_{0\lambda}$ is the high-temperature ($T\geq T_\mathrm{N}$) phonon frequency the $\lambda_{th}$ phonon mode, and $\bigtriangleup_{\lambda}$ is the spin-phonon coupling parameter derived from the DFT calculations. The temperature dependence of the coupled hybrid modes is shown in {Fig \ref{fig4}(c-d)}. One can see both phonon frequencies increase with increasing temperature which qualitatively agrees with the trend observed in experiments. The spin-phonon coupled hybrid modes in the heterostructure could be attributed to the phonon-resonance phenomenon due to the proximity effect.

\section*{Conclusion}
In conclusion, we report a proximate AFM order in  Bi$_{2}$Te$_{3}$, a topological insulator, by investigating the temperature-dependent Raman spectroscopy of Bi$_{2}$Te$_{3}$ (TI)- FePS$_{3}$ (AFM with $T_\mathrm{N}$ $\sim$ 120 K) stacked vdW heterostructure down to 5 K. Unlike isolated Bi$_{2}$Te$_{3}$, a deviation from the usual phonon anharmonicity in Raman modes corresponding to Bi$_{2}$Te$_{3}$ in the heterostructure was observed at 60 K and can be correlated to antiferromagnetic proximity induced spin-phonon coupling. The strength of spin-phonon coupling decreases with the reduction of thickness of Bi$_{2}$Te$_{3}$ and FePS$_{3}$ in heterostructure, but the characteristic temperature for spin-phonon coupling remains invariant ($\sim$ 60 K). The robust spin-phonon coupling in Bi$_{2}$Te$_{3}$  could be destroyed by placing hBN in between Bi$_{2}$Te$_{3}$ and FePS$_{3}$ in the heterostructure. Also, a reduction of $T_\mathrm{N}$ to 65 K of hybrid FePS$_{3}$ was identified, possibly due to the interfacial strain which leads to smaller Fe-S-Fe bond angles as corroborated by DFT calculations. The current study on the spatial variation of spin-phonon coupling at the interface of vdW magnet-TI heterostructure may be crucial for the future spin logic devices.


\textbf{Methods}\\
\textbf{Sample preparation}\\
Binary  Bi$_{2}$Te$_{3} $  single crystal was grown through solid-state reaction method by using box furnace. Bismuth powder (Alfa Aesar, purity 99.999\% ) , Tellurium powder (Alfa Aesar , purity 99.999\% ) were mixed in proper stoichiometric ratio and sealed into  a quartz ampoule under vacuum($\approx$10$^{-5}$ mbar). The ampoule was then placed at the centre of the furnace and heated at 700$^{\circ}$ C for seven days \cite{koyano2012single}.\\
\textbf{Computational Details}\\ 
Density functional theory (DFT) calculations have been performed by using a plane-wave basis set with a kinetic energy cutoff of 400 eV and projector augmented-wave~\cite{bl1994hl,kresse1999ultrasoft} potentials as implemented in the Vienna Ab initio Simulation Package (VASP)~\cite{PhysRevB.54.11169, kresse1996efficiency}. For the exchange-correlation functional, the Perdew-Burke-Ernzerhof (PBE)~\cite{perdew1996generalized} version of the generalized gradient approximation (GGA) has been used. In our calculations, the spin-polarized case for FePS$_{3}$ and the non-spin-polarized case for Bi$_{2}$Te$_{3}$ have been considered. During structural relaxations, the positions of the ions were relaxed until the Hellman-Feynman forces became less than 0.001 eV/\AA. Correlation effects for Fe $d$ electrons have been incorporated within GGA+$U$~\cite{dudarev1998electron} approach, and an effective on-site Coulomb repulsion $U_{eff}$ = 3 eV, which is within the range of $U$ values used for \cite{U_FePS3,U_MPS3,dey2016orbital} has been considered. In addition, spin-orbit coupling (SOC) was included in our calculations to get the correct electronic band dispersion for Bi$_{2}$Te$_{3}$ and estimate the magnetic anisotropy for FePS$_{3}$. Phonons were calculated using the density functional perturbation theory (DFPT) \cite{baroni2001phonons} as implemented in the PHONOPY \cite{togo2015first}.

The Bi$_{2}$Te$_{3}$-FePS$_{3}$ heterostructure has been constructed within the Quantum ATK framework \cite{smidstrup2019quantumatk}. The surface matching algorithm described in ref ~\cite{stradi2017method} was utilized to obtain low-strain hetero-interfaces. For this heterostructure, the mean absolute strain values on both the monolayer surfaces were calculated as $\sim$0.5\%. The composite supercell was then fully relaxed within the vdW correction method DFT-D3 \cite{grimme2010consistent} until forces on atoms became less than 0.02 eV/\AA. The reciprocal space integration was carried out with a $\Gamma$ centered k-mesh of $8\times8\times2$ for Bi$_{2}$Te$_{3}$ bulk, $12\times12\times1$ for Bi$_{2}$Te$_{3}$ single layer, $12\times8\times1$ for FePS$_{3}$ monolayer, and $2\times2\times1$ for the Bi$_{2}$Te$_{3}$-FePS$_{3}$ heterostructure.

 For Magnon Stiffness Calculation, We employ the interface builder in the QuantumATK package \cite{smidstrup2019quantumatk} to construct a unit cell of the 1ML-FePS$_{3}$/6QL-Bi$_{2}$Te$_{3}$ heterostructure with  $2\times2$ supercell of FePS$_{3}$ and $\sqrt{7}\times\sqrt{7}$ supercell of Bi$_{2}$Te$_{3}$. Spin wave stiffness constant $D$ of 1ML-FePS$_{3}$ /6QL-Bi$_{2}$Te$_{3}$ heterostructure is calculated using Green's function (GF) formalism of Ref. \cite{PRB174402}, as implemented in QuantumATK package \cite{smidstrup2019quantumatk}. The Kohn-Sham Hamiltonian of density functional theory (DFT), as the input of GF formalism, is obtained from noncollinear DFT calculations using the Perdew-Burke-Ernzerhof (PBE) parametrization \cite{PRL3865} of the generalized gradient approximation (GGA) to the exchange-correlation functional, as implemented in QuantumATK package  \cite{smidstrup2019quantumatk}; norm-conserving fully relativistic pseudopotentials of the type PseudoDojo-SO \cite{smidstrup2019quantumatk,VANSETTEN201839} for describing electron-core interactions; and the Pseudojojo (medium) numerical linear combination of atomic orbitals (LCAO) basis set \cite{VANSETTEN201839}. The energy mesh cutoff for the real-space grid is chosen as 101 Hartree, and the k-point grid $6\times6\times1$ is used for the self-consistent calculations. Periodic boundary condition is used for the self-consistent calculations, and a 15\AA \,vacuum is placed on the top of heterostructure in order to remove interaction between the consecutive periodic image.

\textbf{Acknowledgments}\\
The authors would like to thank Dr. Vasant Sathe, Prof. K Sengupta, Dr. Marek Potemski, Dr. Cl{\'e}ment Faugeras, Prof. Achintya Singha, Mr. Somsubhra Ghosh, Dr. Mintu Mondal, and Dr. Devajyoti Mukherjee for fruitful discussion. SMaity and TK are grateful to DST-INSPIRE for their fellowships. D.D. and L.Y. gratefully acknowledge the U.S. DOE, Office of Science, Office of Basic Energy Sciences, under Award No. DE-SC0021127 for financial assistance and Advanced Computing Group of the University of Maine System for providing computational resources for this work.  SMasanta is grateful to Council of Scientific \& Industrial Research (CSIR), New Delhi, for the financial support through the award of NET-SRF (File No: 09/015(0531)/2018-EMR-I). The authors are thankful to the facilities at UGC-DAE-CSR-Indore. Magneto-Raman scattering at low temperature were performed at LNCMI, European Magnetic Field Laboratory at Grenoble under the project GSC08-119. SMaity would like to thank Ms. Diana Vaclavkova, Ms. Anushree Dey, Mr. Sanjib Naskar, Mr. Rahul Paramanik and Mr. Soumik Das. MP and BD is grateful to IACS for the fellowship. SD acknowledges the financial support from DST-SERB grant No. ECR/2017/002037, SCP/2022/000411 and CRG/2021/004334. SD also acknowledges support from the Central Scientific Service (CSS) and the Technical Research Centre (TRC), IACS, Kolkata. 
\bibliography{Manuscript}

\end {document}


\title{Supplementary Information: Manipulating Spin-Lattice Coupling in  Layered Magnetic Topological Insulator Heterostructure \textit{via} Interface Engineering}
\author{Sujan Maity}
\affiliation{School of Physical Sciences, Indian Association for the Cultivation of Science, 2A \&  2B Raja S. C. Mullick Road, Jadavpur, Kolkata - 700032, India}
\author{Dibyendu Dey}
\affiliation{Department of Physics and Astronomy, University of Maine, Orono, ME 04469, USA}

\author{Anudeepa Ghosh}

\affiliation{School of Physical Sciences, Indian Association for the Cultivation of Science, 2A \& 2B Raja S. C. Mullick Road, Jadavpur, Kolkata - 700032, India}

\author{Suvadip Masanta}
\affiliation{Bose Institute, Department of Physics
Main Campus 
93/1, A. P. C. Road
Kolkata - 700 009, India}



\author{Binoy Krishna De}
\affiliation{UGC-DAE Consortium for Scientific Research,
Indore Centre,
University Campus, Khandwa Road, Indore 452001}

\author{Hemant Singh Kunwar}
\affiliation{UGC-DAE Consortium for Scientific Research,
Indore Centre,
University Campus, Khandwa Road, Indore 452001}

\author{Bikash Das}
\affiliation{School of Physical Sciences, Indian Association for the Cultivation of Science, 2A \& 2B Raja S. C. Mullick Road, Jadavpur, Kolkata - 700032, India}
\author{Tanima Kundu}
\affiliation{School of Physical Sciences, Indian Association for the Cultivation of Science, 2A \& 2B Raja S. C. Mullick Road, Jadavpur, Kolkata - 700032, India}

\author{Mainak Palit}
\affiliation{School of Physical Sciences, Indian Association for the Cultivation of Science, 2A \& 2B Raja S. C. Mullick Road, Jadavpur, Kolkata - 700032, India}

\author{Satyabrata Bera}
\affiliation{School of Physical Sciences, Indian Association for the Cultivation of Science, 2A \& 2B Raja S. C. Mullick Road, Jadavpur, Kolkata - 700032, India}

\author{Kapildeb Dolui}

\affiliation{Lomare Technolgies Limited, 6 London Street, London EC3R 7LP, United Kingdom}

\author{Kenji Watanabe}
\affiliation{Research Center for Functional Materials,National Institute for Materials Science, Tsukuba 305-0044, Japan}
\author{Takashi Taniguchi}
\affiliation{International Center for Materials Nanoarchitectonics, National Institute for Materials Science,Tsukuba 305-0044, Japan}

\author{Liping Yu }
\affiliation{Department of Physics and Astronomy, University of Maine, Orono, ME 04469, USA}

\author{A Taraphder}
\affiliation{Department of Physics, Indian Institute of Technology Kharagpur, W.B. 721302, India}

\author{Subhadeep Datta*}
\affiliation{School of Physical Sciences, Indian Association for the Cultivation of Science, 2A \& 2B Raja S. C. Mullick Road, Jadavpur, Kolkata - 700032, India}
\email{sspsdd@iacs.res.in}
\maketitle

\begin{figure}[H]
\centerline{\includegraphics[scale=0.6, clip]{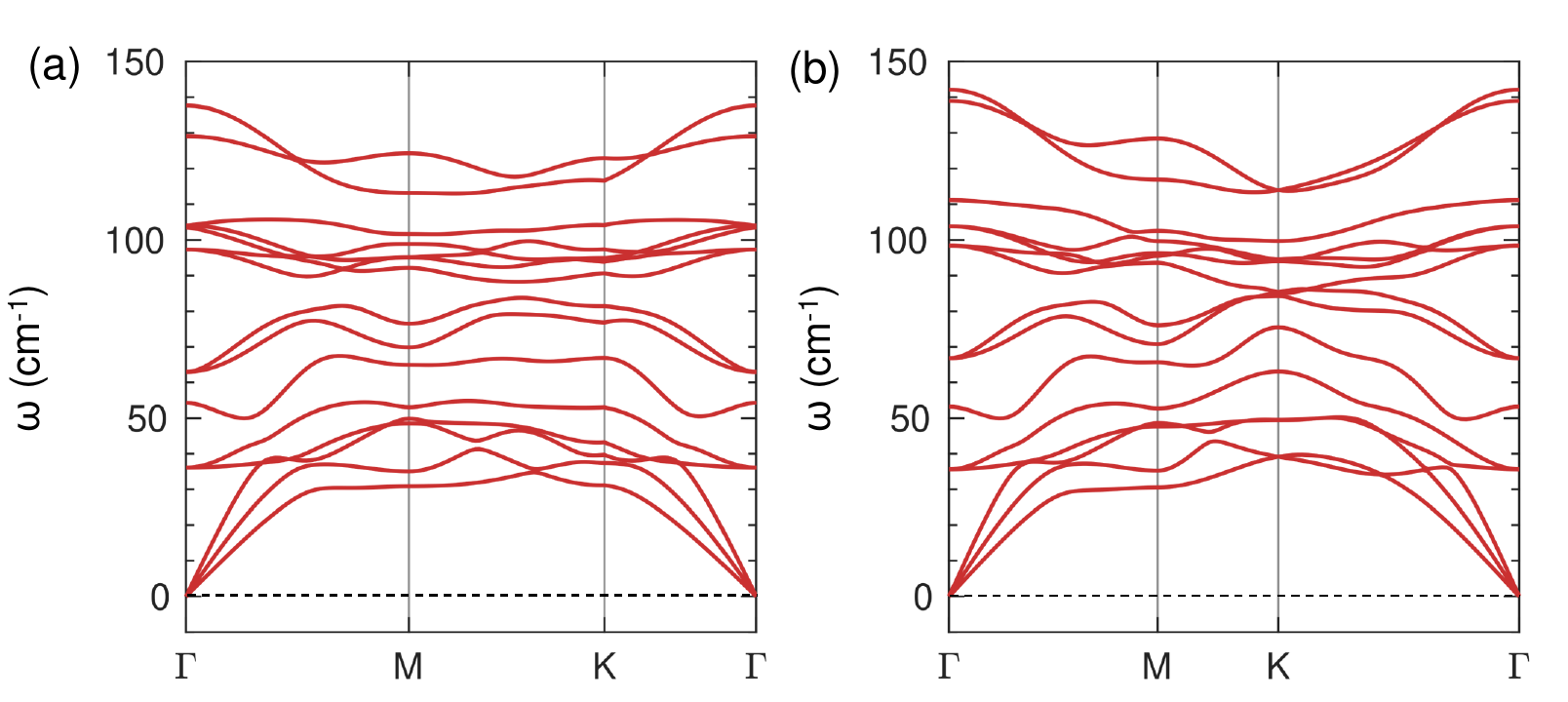}}
\caption{Phonon dispersion of Bi$_{2}$Te$_{3}$ (a) bulk and (b) monolayer structures, respectively. The phonon dispersion of Bi$_{2}$Te$_{3}$ monolayer agree well with its bulk counterpart. No imaginary frequencies have been observed in phonon dispersion, suggesting that this structure is dynamically stable.}
\end{figure}

\begin{figure}[H]
\centerline{\includegraphics[scale=0.6, clip]{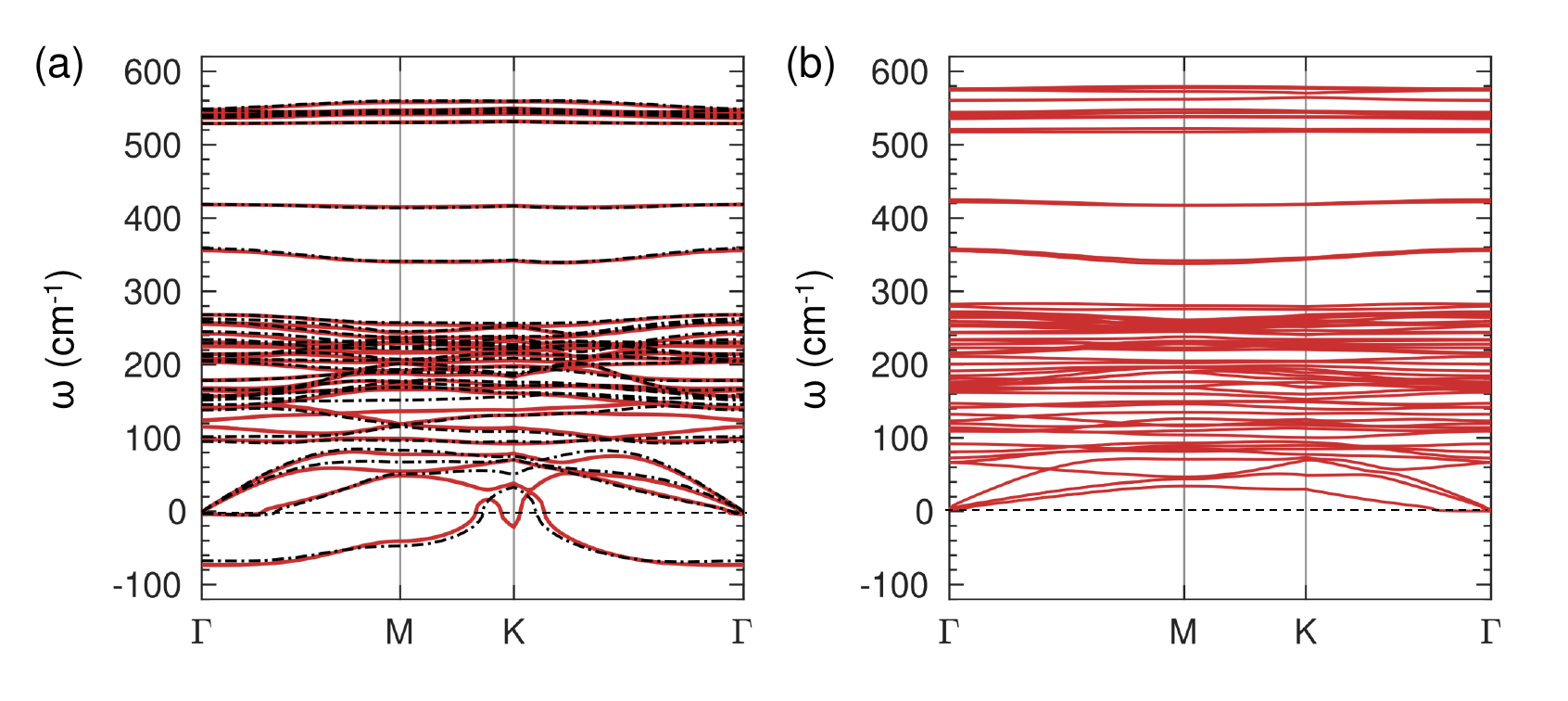}}
\caption{(a) Phonon dispersion of FePS$_{3}$ monolayer for z-AFM (red solid lines) and FM (black dotted lines) spin configurations in the experimentally reported point-group structure. (b) Phonon dispersion of distorted FePS$_{3}$ monolayer for z-AFM spin configuration indicate the dynamical stability due to the absence of imaginary modes}
\end{figure}

\begin{figure}[H]
\centerline{\includegraphics[scale=0.6, clip]{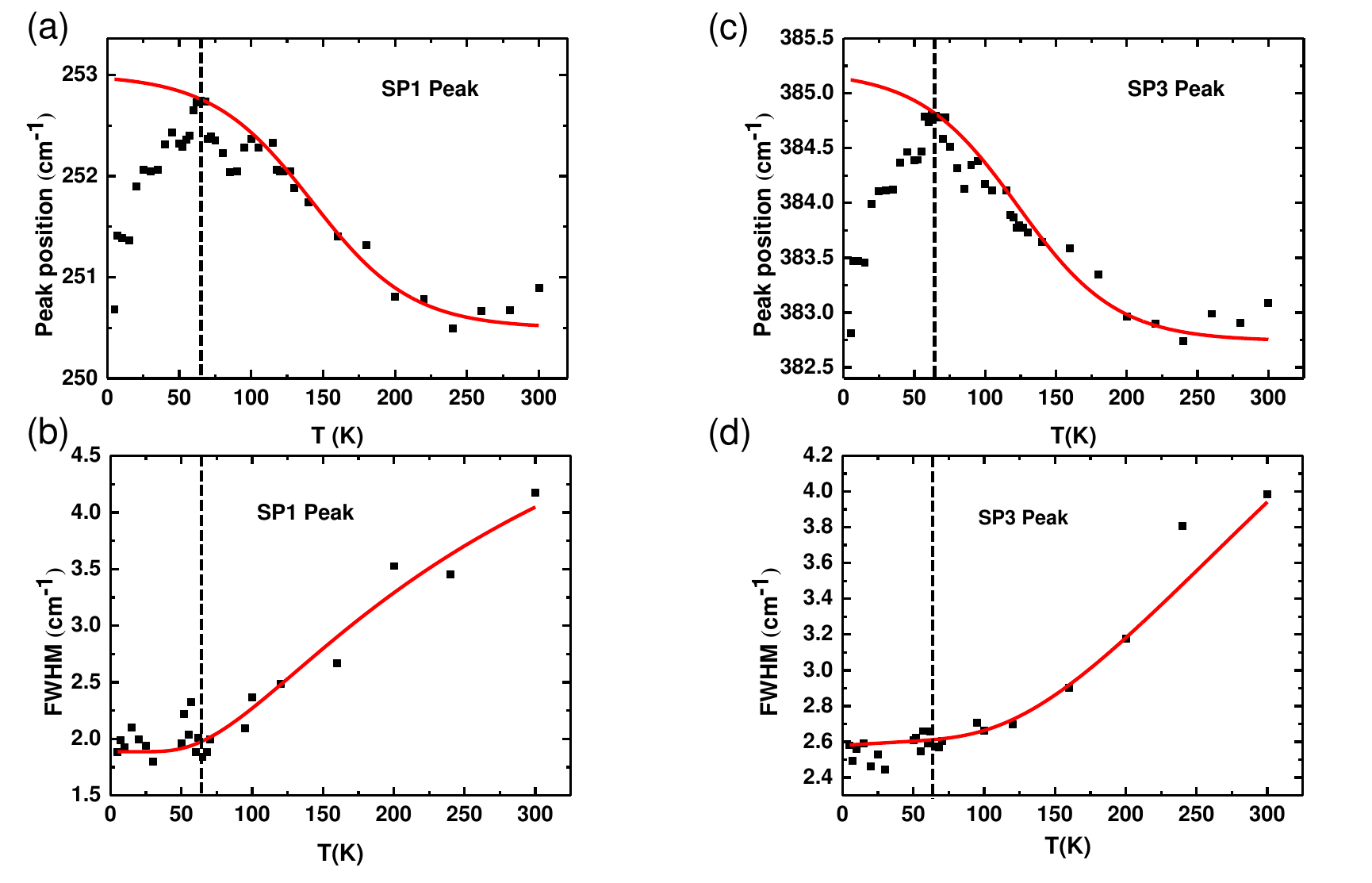}}
\caption{Variation of Peak position and linewidth of FePS$_{3}$ Raman modes connected to heterostructure  with temperature. (a),(b) SP1 mode; (c), (d)  SP3 mode; All spin-phonon peaks show phonon anomaly at around 65 K in phonon frequency.  Linewidths of all spin-phonon mode of FePS$_{3}$ are fit with 3-phonon anharmonic decay (AD) model. }
 \end{figure}
\begin{figure}[H]
\centerline{\includegraphics[scale=0.6, clip]{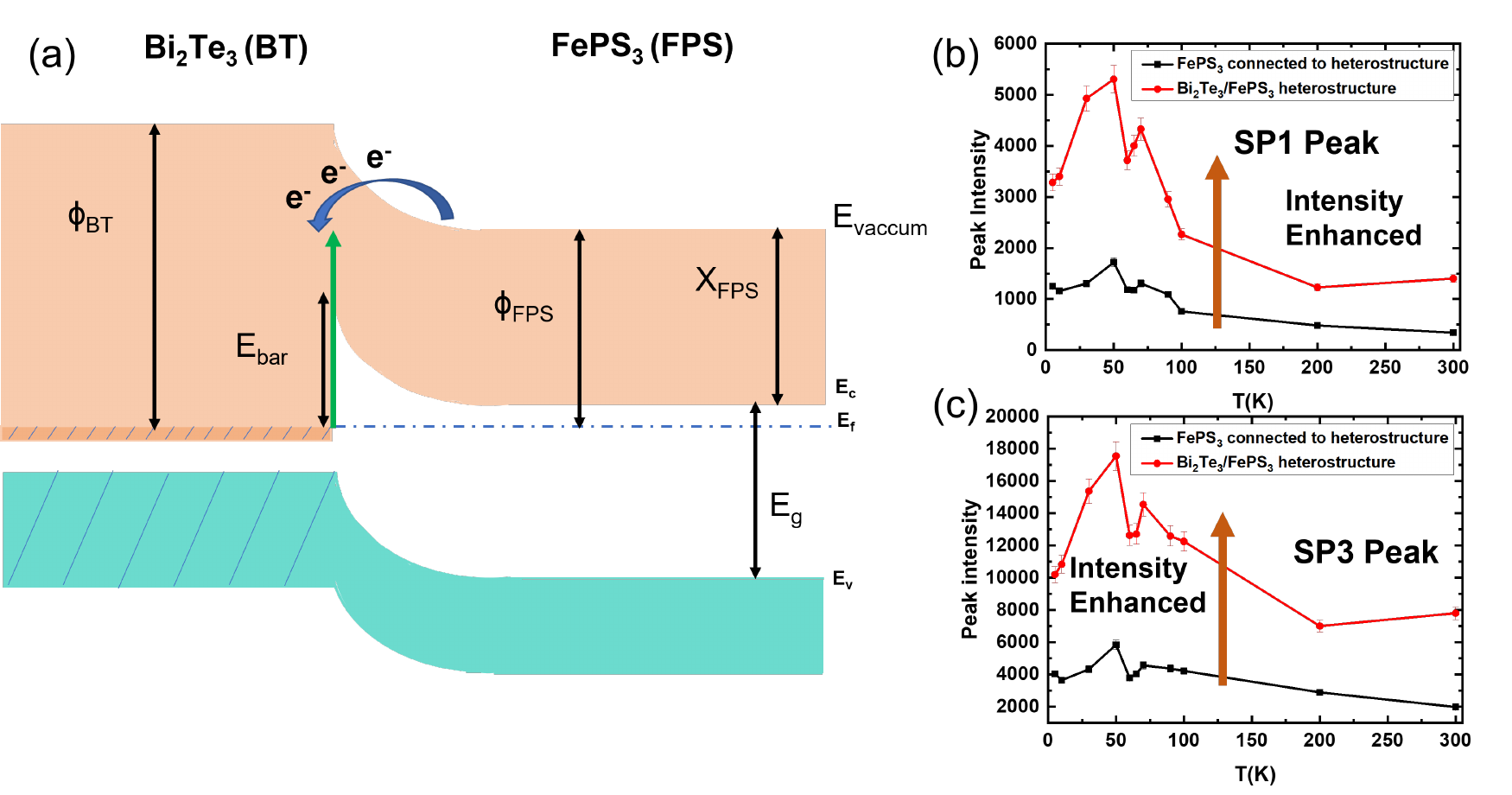}}
\caption{(a) Schematic diagram of the band bending that occurs at the interface between the metallic Bi$_{2}$Te$_{3}$ and the semiconducting FePS$_{3}$. The work function of Bi$_{2}$Te$_{3}$ ($\phi_{Bi_{2}Te_{3}}$ = 5.30 eV) is larger than that of FePS$_{3}$ ($\phi_{FePS_{3}}$ = 4.23 eV) which leads to the formation of a Schottky barrier of height $\phi_{Bi_{2}Te_{3}}$ - $\chi_{FePS_{3}}$ = 1.17 eV at the interface. To balance out the chemical potential electrons move from the FePS$_{3}$ into the Bi$_{2}$Te$_{3}$. (b), (c) Intensity of Raman peaks of FePS$_{3}$ are enhanced due to Bi$_{2}$Te$_{3}$ underneath compare to isolated FePS$_{3}$ Raman modes. In Bi$_{2}$Te$_{3}$-FePS$_{3}$ heterostructure, barrier height (E$_{bar}$) (1.17 eV)$ <$ Raman excitation (denoted by green arrow) (2.62 eV). The available transitions for Raman scattering will then involve real energy levels, drastically enhancing the intensity}
\end{figure}
\begin{figure}[H]
\centerline{\includegraphics[scale=0.6, clip]{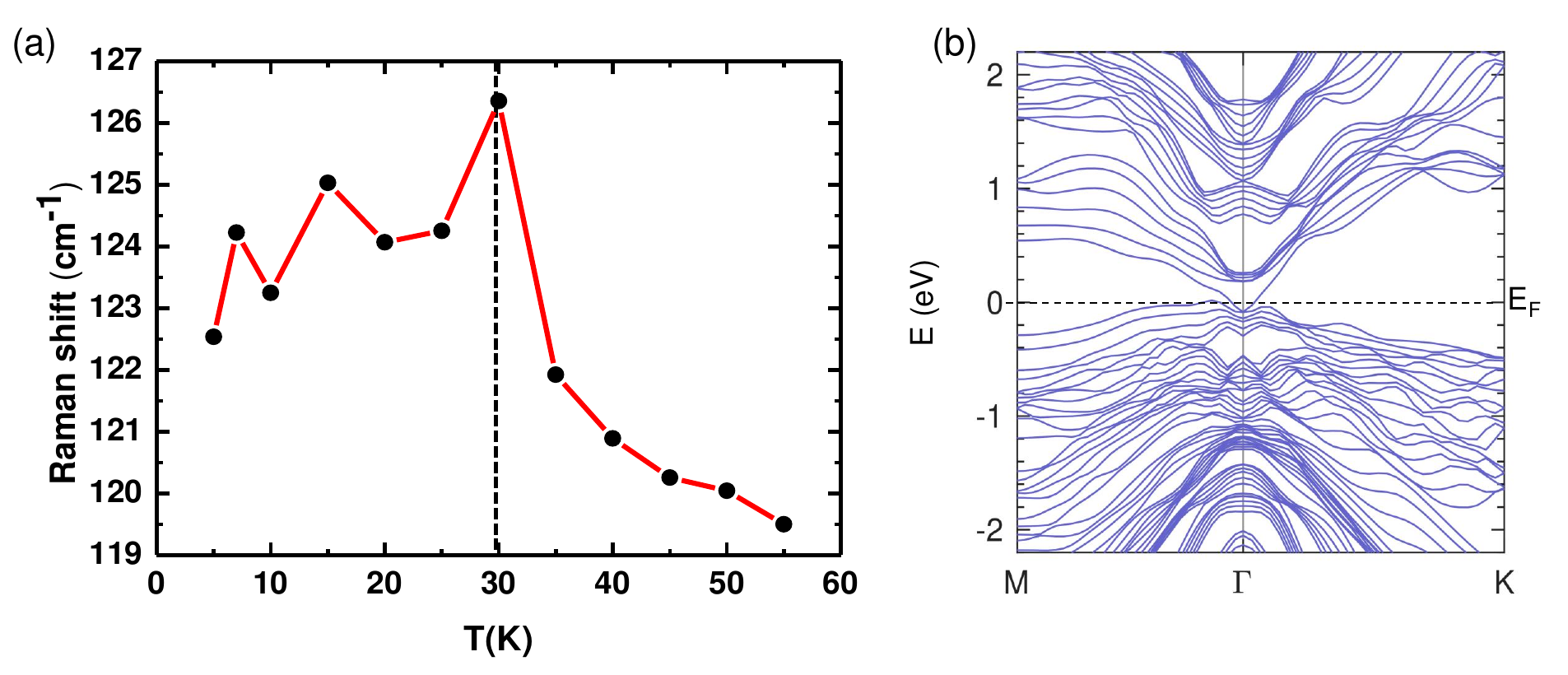}}
\caption{(a) Magnon mode of FePS$_{3}$ connected to heterostructure. One-magnon mode at 120 cm$^{-1}$ at a temperature ($T_{M} \sim 60 K$) was observed in FePS$_{3}$. Softening of this magnon mode was observed with the temperature and the softening temperature is 30K in case of FePS$_{3}$. (b) Electronic structure of 6 quintuple layers of Bi$_{2}$Te$_{3}$. Metallic surface states start appearing at a thickness ($\sim$ 6 nm) of 6 quintuple layers.}
\end{figure}

\begin{figure}[H]
\centerline{\includegraphics[scale=0.6, clip]{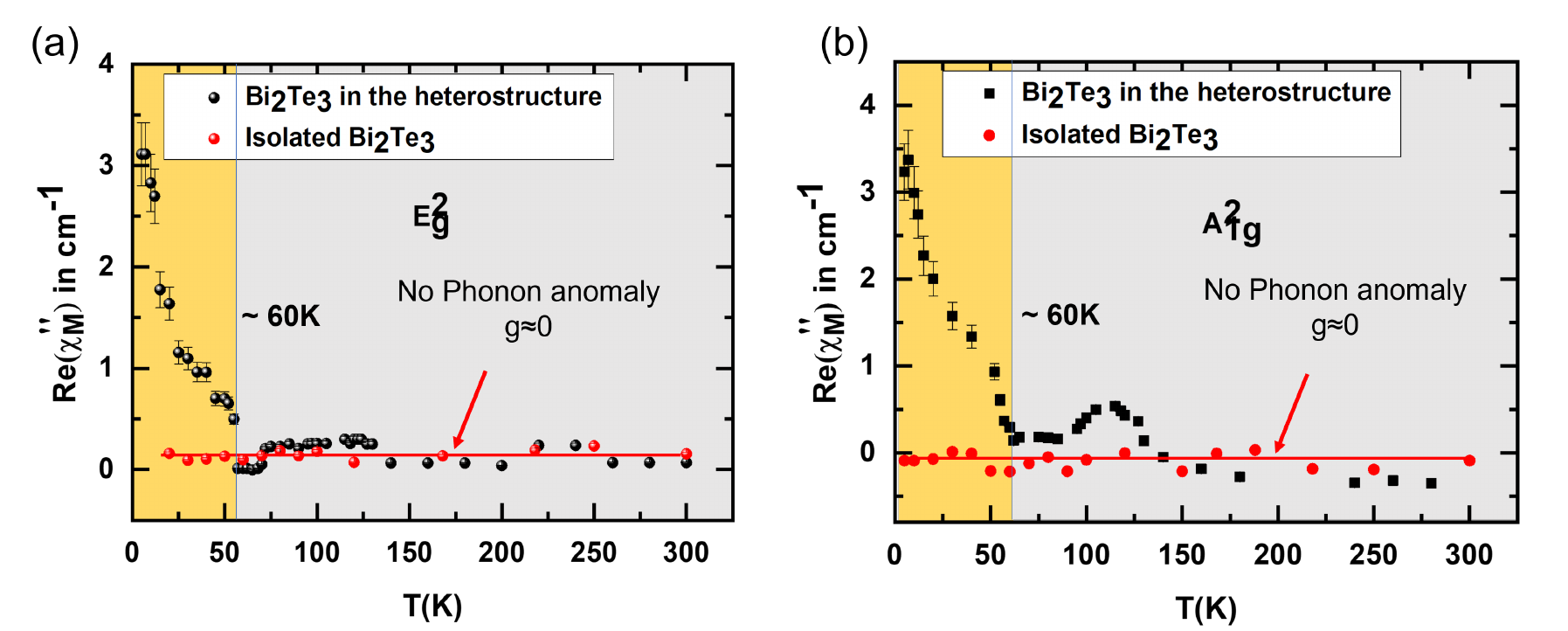}}
\caption{ Real part of Magnetic Susceptibility of (a) In-plane E$_{g}^{2}$ mode; (b) Out-of-plane  A$_{1g}^{2} $ mode of Bi$_{2}$Te$_{3}$ in Bi$_{2}$Te$_{3}$/FePS$_{3}$ heterostructure ;  No such change in spin-phonon coupling constant (g) was observed with temperature in both Raman modes of isolated Bi$_{2}$Te$_{3}$ due to pure phononic behaviour in the whole temperature range. In case of Bi$_{2}$Te$_{3}$, in the heterostructure, slope of the curve changes around 60K and this is the reflection of proximity induced spin-phonon coupling of Bi$_{2}$Te$_{3}$ , occurs at/around 60K.}
\end{figure}

\begin{figure}[H]
\centerline{\includegraphics[scale=0.6, clip]{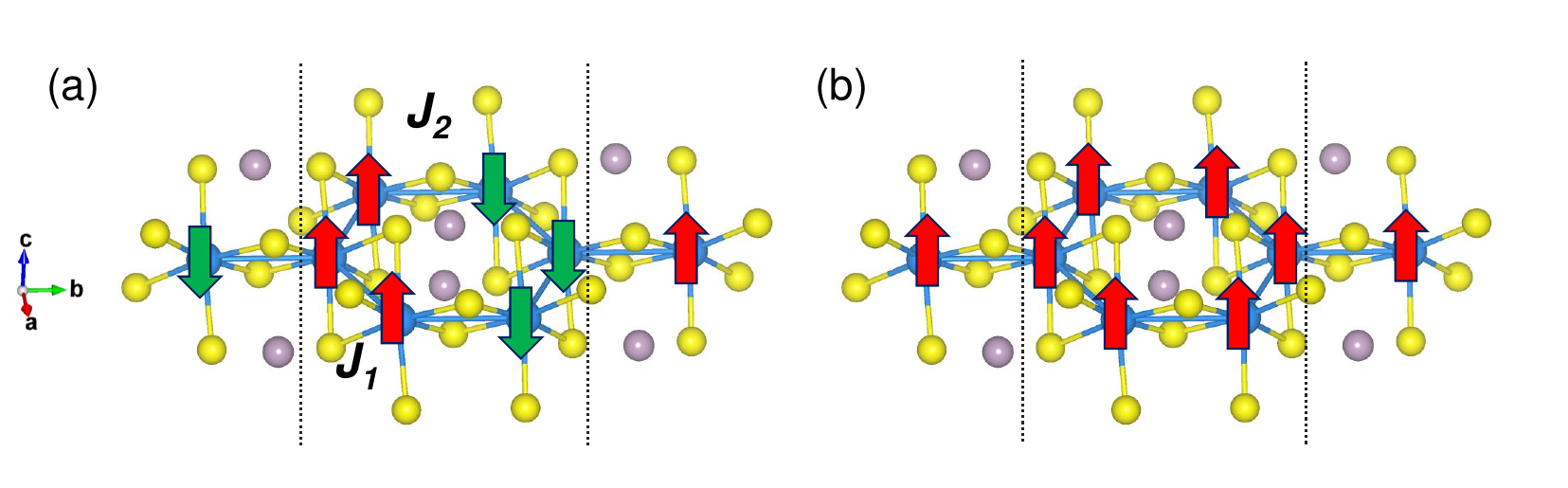}}
\caption{(a) zig-zag AFM and (b) FM spin configurations of FePS$_{3}$ monolayer that are used to evaluate Fe-Fe AFM exchange.}
\end{figure}
By mapping the DFT total energy of z-AFM (E$_1$) and FM spin configurations (E$_2$) to the Heisenberg spin model, $H=J$ $\displaystyle\sum _{i>j}S_{i}\cdot S_{j}$, we obtain the following equations- 
\begin{equation}
E_{1}= 4J_{1}S^{2}-2J_{2}S^{2}+\epsilon\\
\end{equation}
\begin{equation}
    E_{2}=4J_{1}S^{2}+2J_{2}S^{2}+\epsilon\\
\end{equation}
\begin{equation}
   J_{2}=\frac{(E_{2}-E_{1})}{4S^{2}}\\
\end{equation}
where, $J_{1}$ is the FM exchange along the a-axis and $J_{2}$ is the AFM exchange along the b-axis. Stronger $J_{2}$ indicates higher T$_N$, whereas weaker $J_{2}$ indicates lower T$_N$. \\
\begin{figure}[H]
\centerline{\includegraphics[scale=0.5, clip]{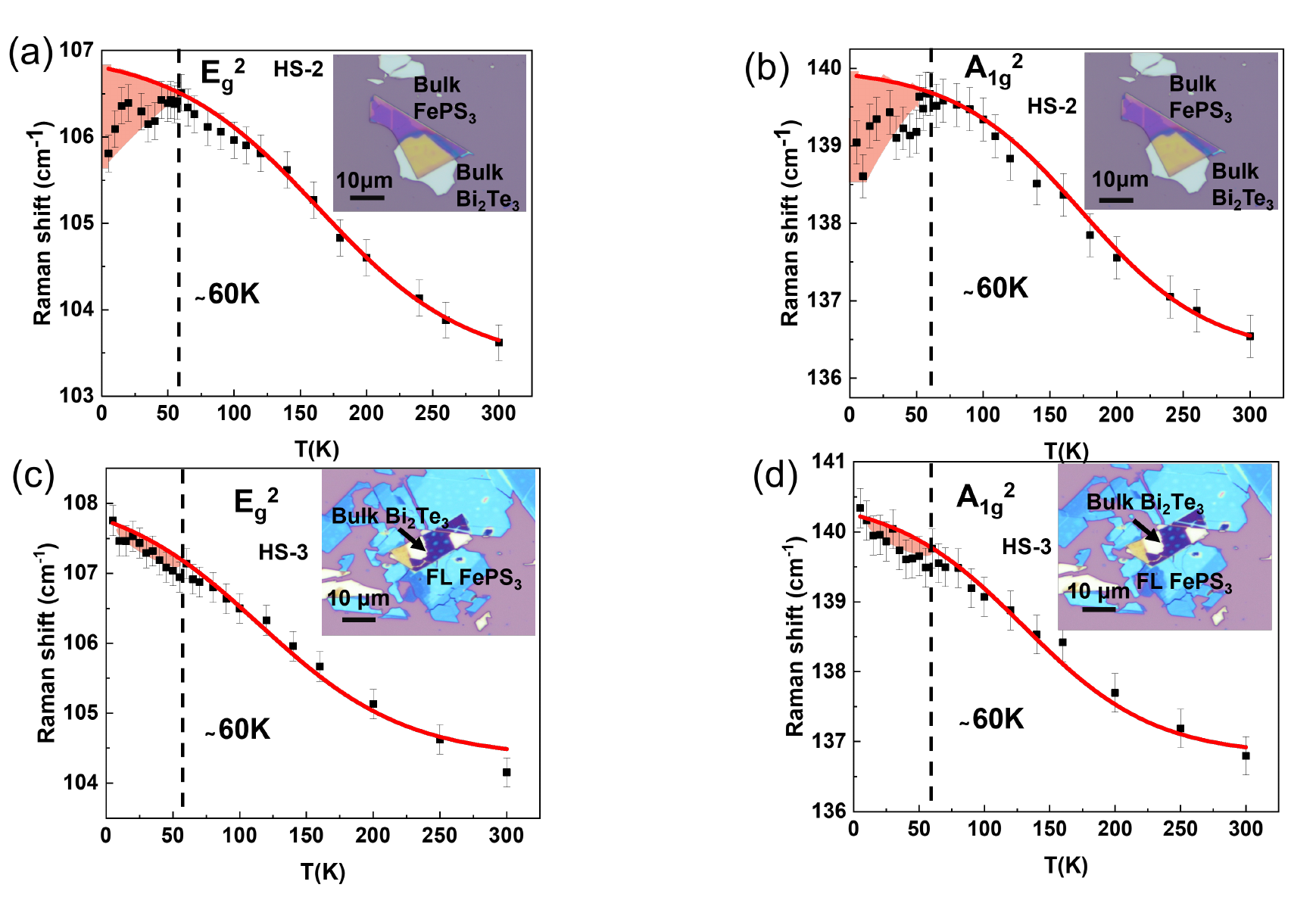}}
\caption{Thickness dependent Raman studies on Bi$_{2}$Te$_{3}$/FePS$_{3}$ heterostructure with temperature.(a) In-plane phonon mode (E$_{g}^{2}$) and (b) Out-of-plane phonon mode (A$_{1g}^{2}$) of Bi$_{2}$Te$_{3}$ in bulk Bi$_{2}$Te$_{3}$ / bulk FePS$_{3}$ heterostructure (HS-2) and, (c) In-plane phonon mode (E$_{g}^{2}$) and (d) Out-of-plane phonon mode (A$_{1g}^{2}$) of Bi$_{2}$Te$_{3}$ in bulk Bi$_{2}$Te$_{3}$/Few layer FePS$_{3}$ heterostructure (HS-3). Insets are showing the optical microscopy images of HS-2 and HS-3. The deviation from anharmonic behaviour, $\Delta\omega$ (related to square of magnetization) values for HS-2 and HS-3 heterostructures, are 0.69 and 0.57 respectively for both phonon modes of Bi$_{2}$Te$_{3}$. Strength of
the spin-phonon coupling decreases but phonon anomaly in both Raman modes of Bi$_{2}$Te$_{3}$ remains invariant ($\sim$ 60K) with reduction of thickness of the heterostructures. }
\end{figure}
\begin{figure}[H]
\centerline{\includegraphics[scale= 0.5, clip]{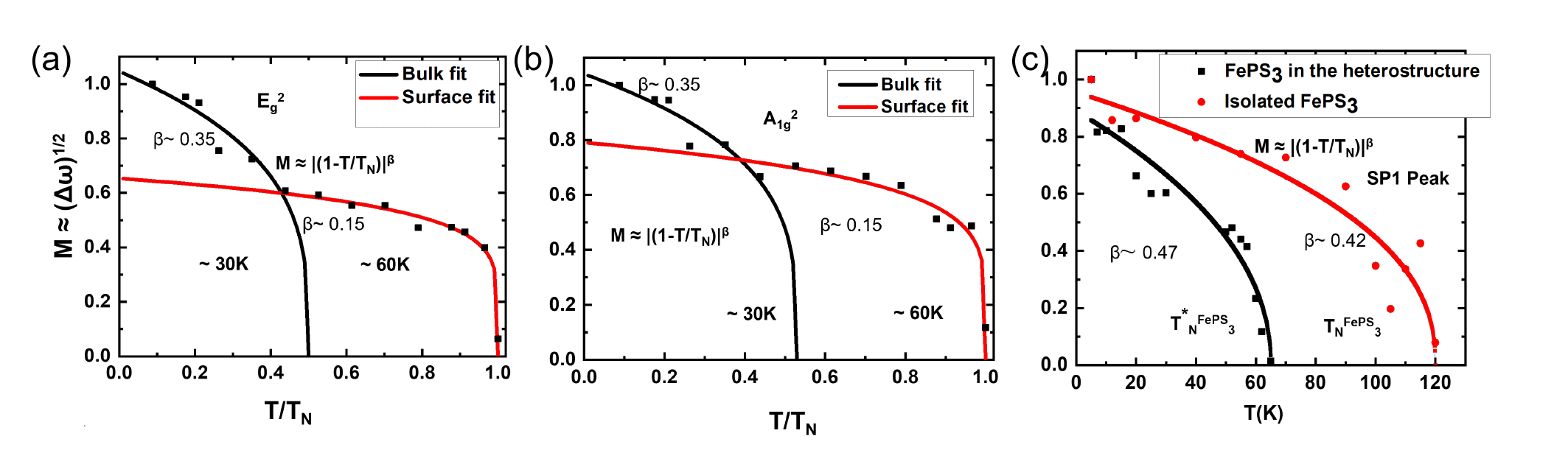}}
\caption{Antiferromagnetic order parameter of (a) In plane E$_{g}^{2}$ mode; (b) out of plane A$_{1g}^{2}$ mode of Bi$_{2}$Te$_{3}$ in the heterostructure ; (c) SP1 peak for two different configurations, FePS$_{3}$ in the heterostructure ( $T_\mathrm{N}^{*}$) and isolated FePS$_{3}$  ($T_\mathrm{N}$) respectively as a function of temperature. Data points are fit with $M\approx |(1-\frac{T}{T_\mathrm{N}})|^{\beta} $ equation. $\beta$ is the critical exponent. The $\beta$ value at 60K (0.15) for surface magnetism corroborates the value of 2D Ising model. The $\beta$ value at 30K (0.35) corresponds to 3D Heisenberg model, responsible for bulk contribution. The $\beta$ values for the FePS$_{3}$ [Fig (c)] are close to mean field value. \label{fig5} }
\end{figure}
\begin{figure}[H]
\centerline{\includegraphics[scale=0.5, clip]{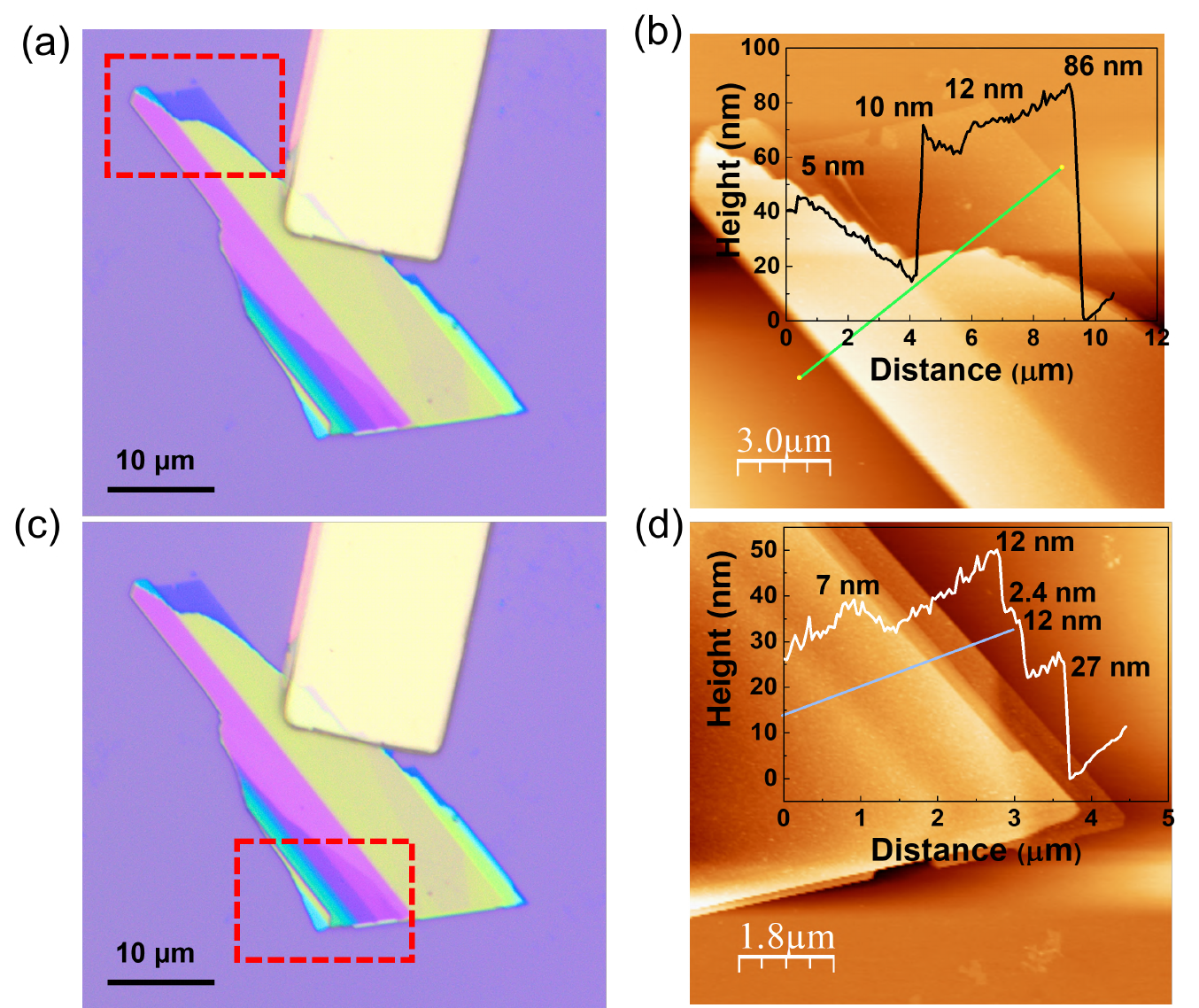}}
\caption{(a), (c) Optical microscopy image of FePS$_{3}$ flake 1. (b), (d) Atomic force microscopy images of the highlighted area of the flake 1. Insets of (b), (d) show the height profile of the flake. }
\end{figure}
\begin{figure}[H]
\centerline{\includegraphics[scale=0.5, clip]{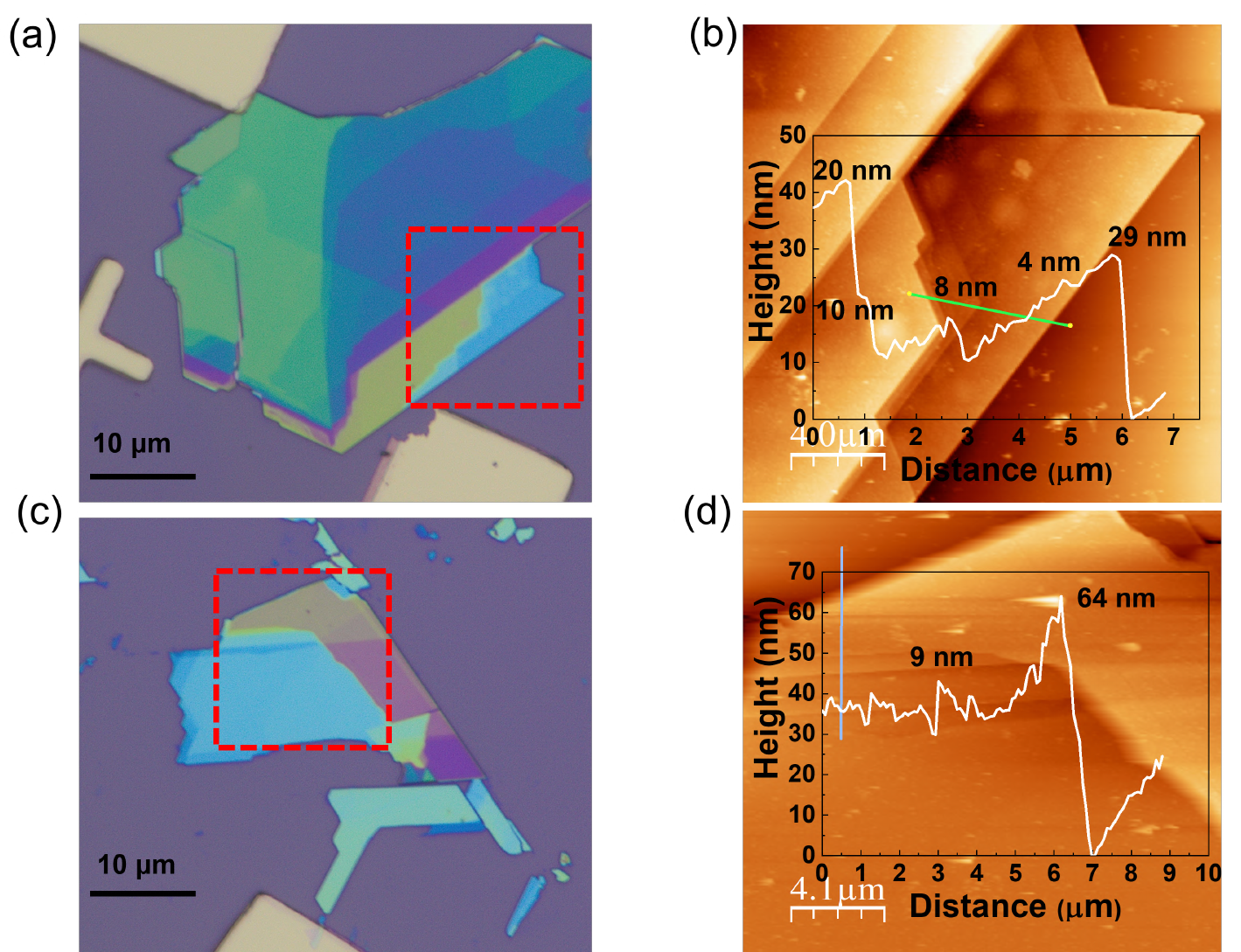}}
\caption{(a), (c) Optical microscopy image of FePS$_{3}$ flake 2 and flake 3. (b), (d) Atomic force microscopy images of the highlighted area of the flake 2 and 3. Insets of (b), (d) show the height profile of the flakes. }
\end{figure}
\begin{figure}[H]
\centerline{\includegraphics[scale=0.5, clip]{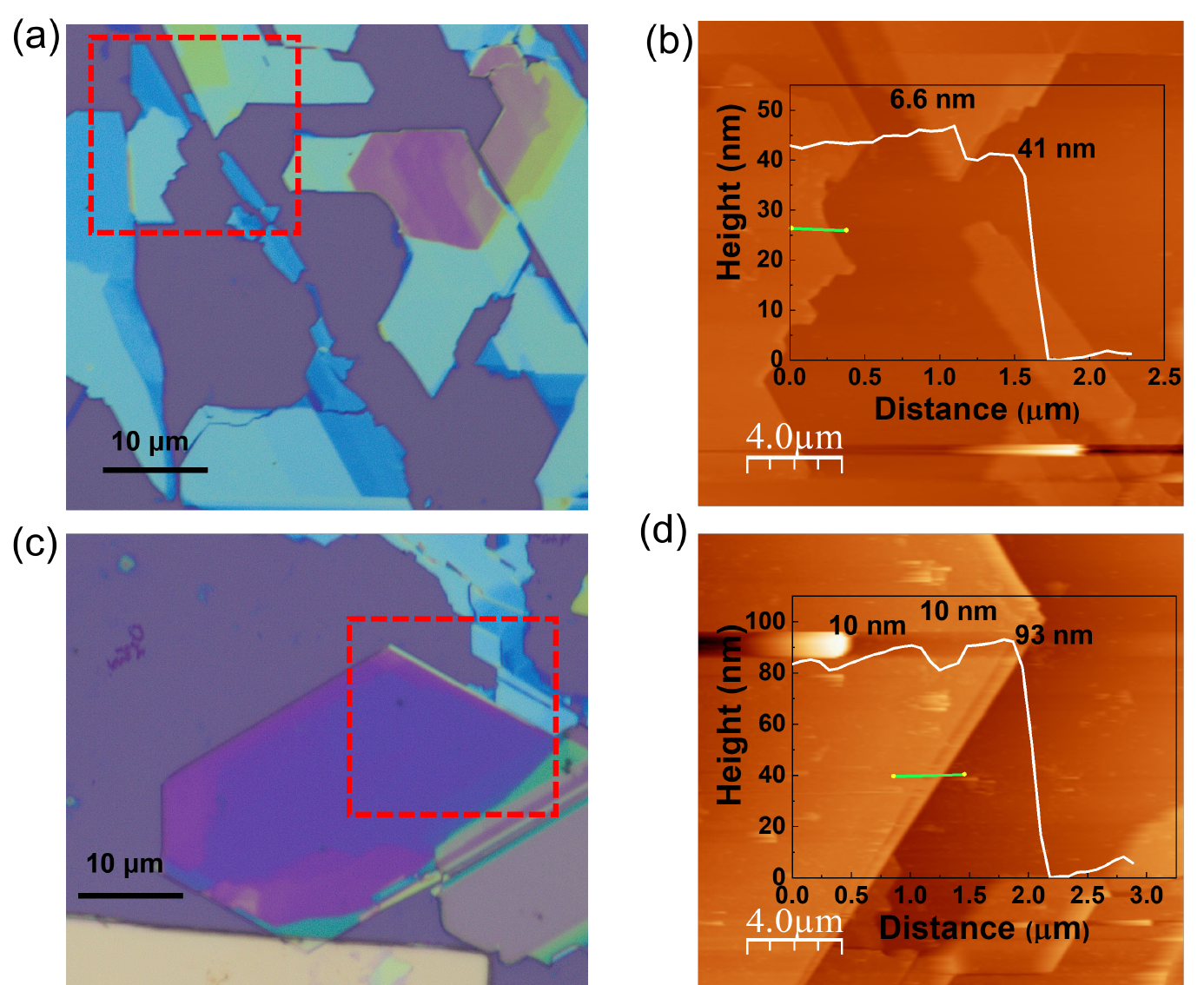}}
\caption{(a), (c) Optical microscopy image of FePS$_{3}$ flake 4 and flake 5. (b), (d) Atomic force microscopy images of the highlighted area of the flake 4 and flake 5. Insets of (b), (d) show the height profile
of the flakes. }
\end{figure}
\begin{figure}[H]
\centerline{\includegraphics[scale=0.5, clip]{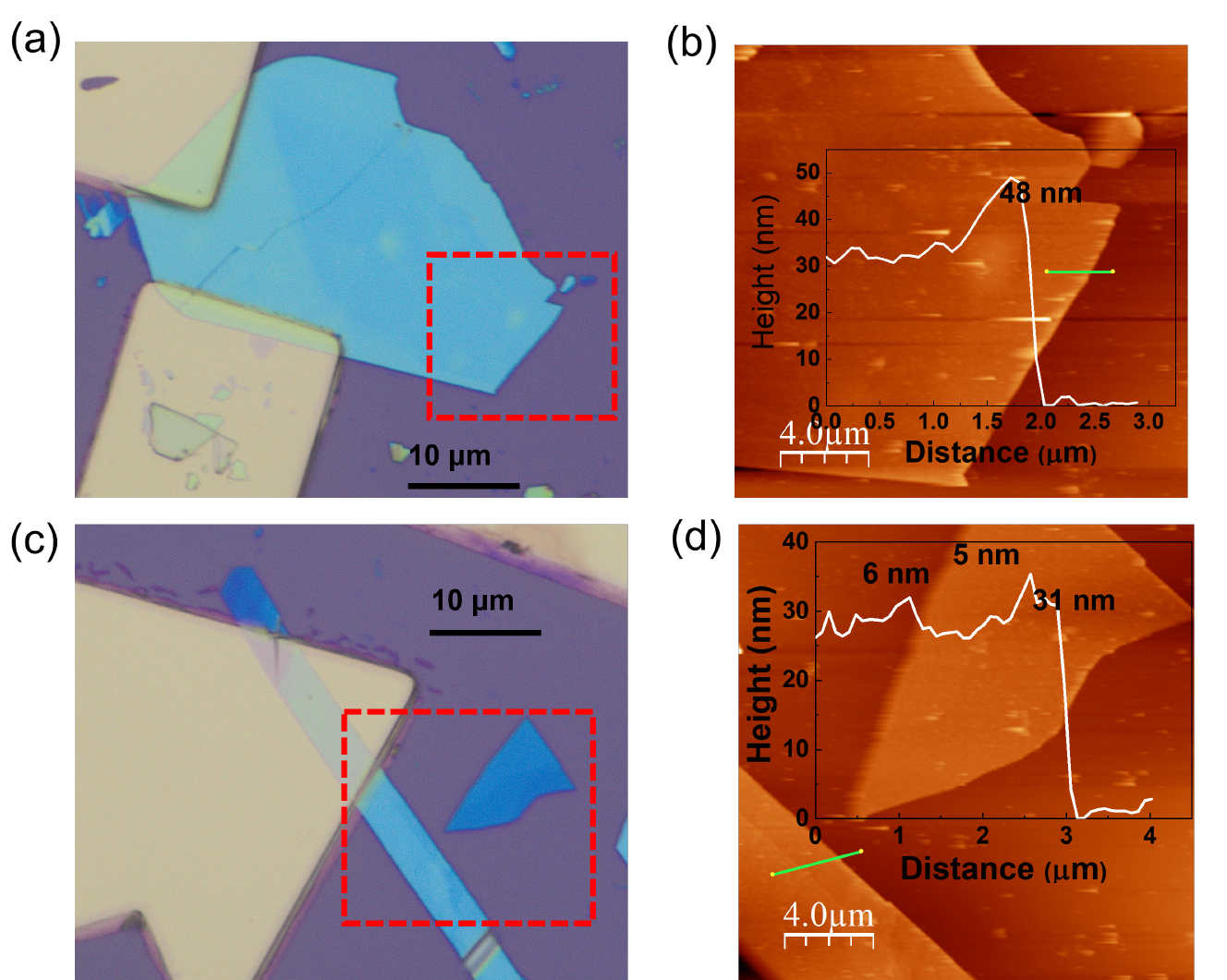}}
\caption{(a), (c) Optical microscopy image of FePS$_{3}$ flake 6 and flake 7. (b), (d) Atomic force microscopy images of the highlighted area of the flake 6 and flake 7. Insets of (b), (d) show the height profile
of the flakes.}
\end{figure}
\begin{figure}[H]
\centerline{\includegraphics[scale=0.5, clip]{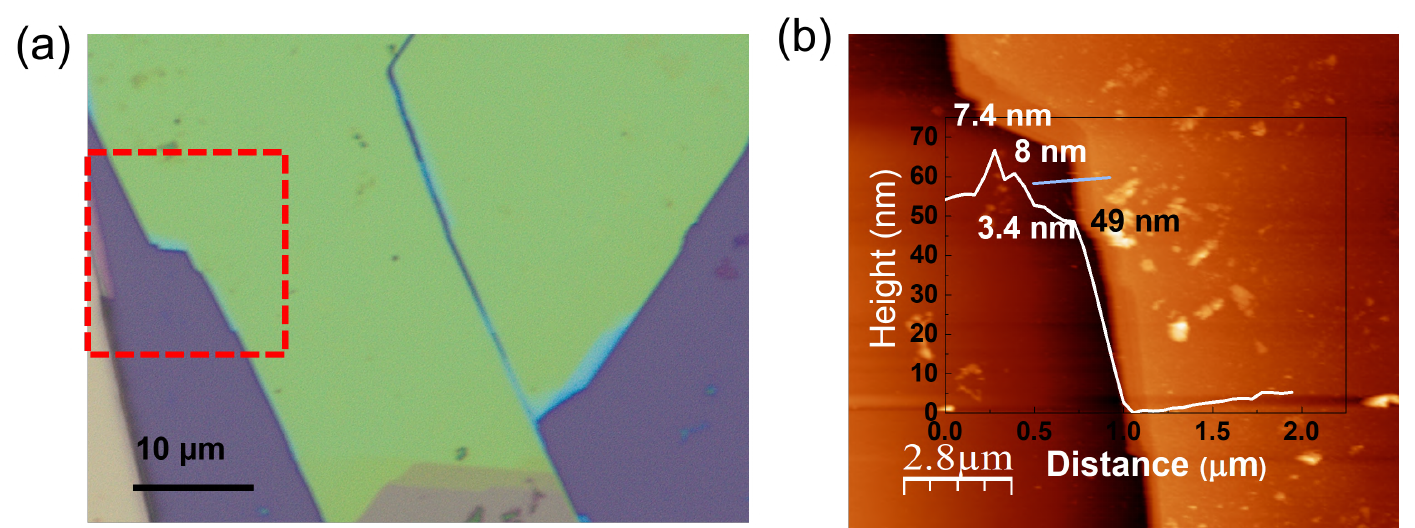}}
\caption{(a) Optical microscopy image of FePS$_{3}$ flake 8. (b) Atomic force microscopy image of the highlighted area of the flake 8. Inset of (b) shows the height profile of the flake.}
\end{figure}
\begin{figure}[H]
\centerline{\includegraphics[scale=0.5, clip]{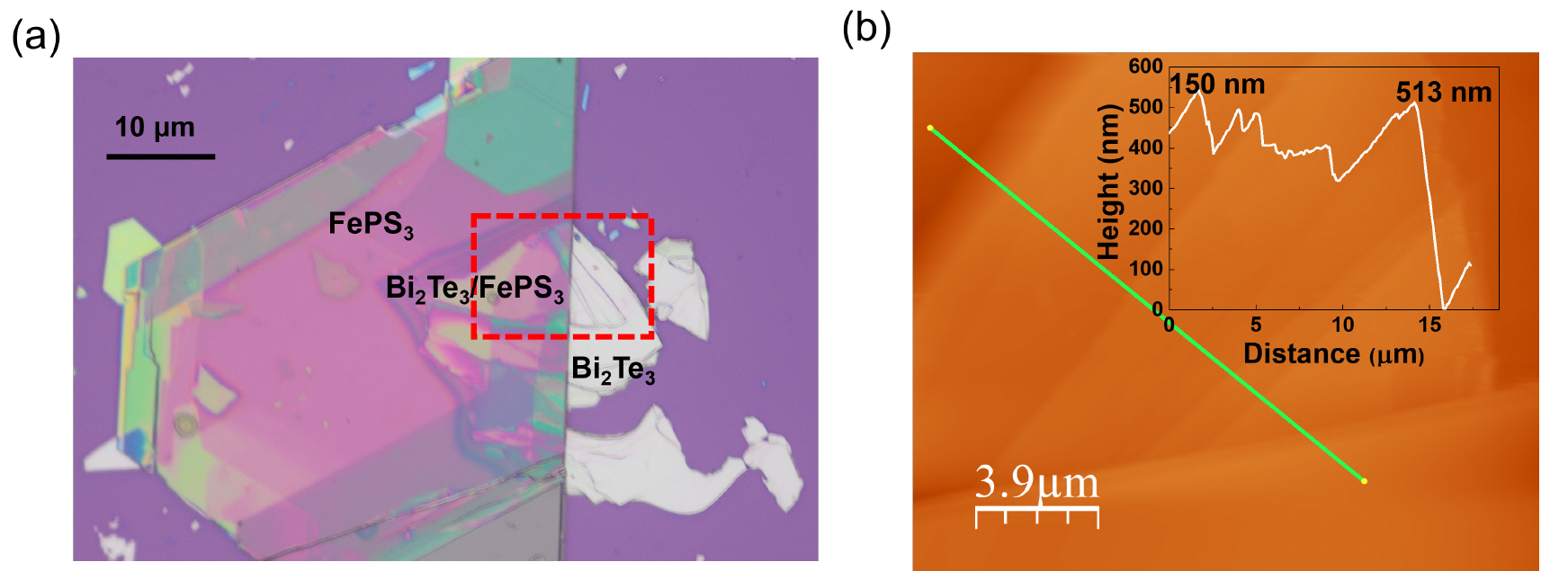}}
\caption{(a) Optical microscopy image of Bi$_{2}$Te$_{3}$/FePS$_{3}$ heterostructure (HS-1). (b) Atomic force microscopy image of Bi$_{2}$Te$_{3}$/FePS$_{3}$ (HS-1). Inset of (b) shows the height profile of the heterostructure.}
\end{figure}
\begin{figure}[H]
\centerline{\includegraphics[scale=0.5, clip]{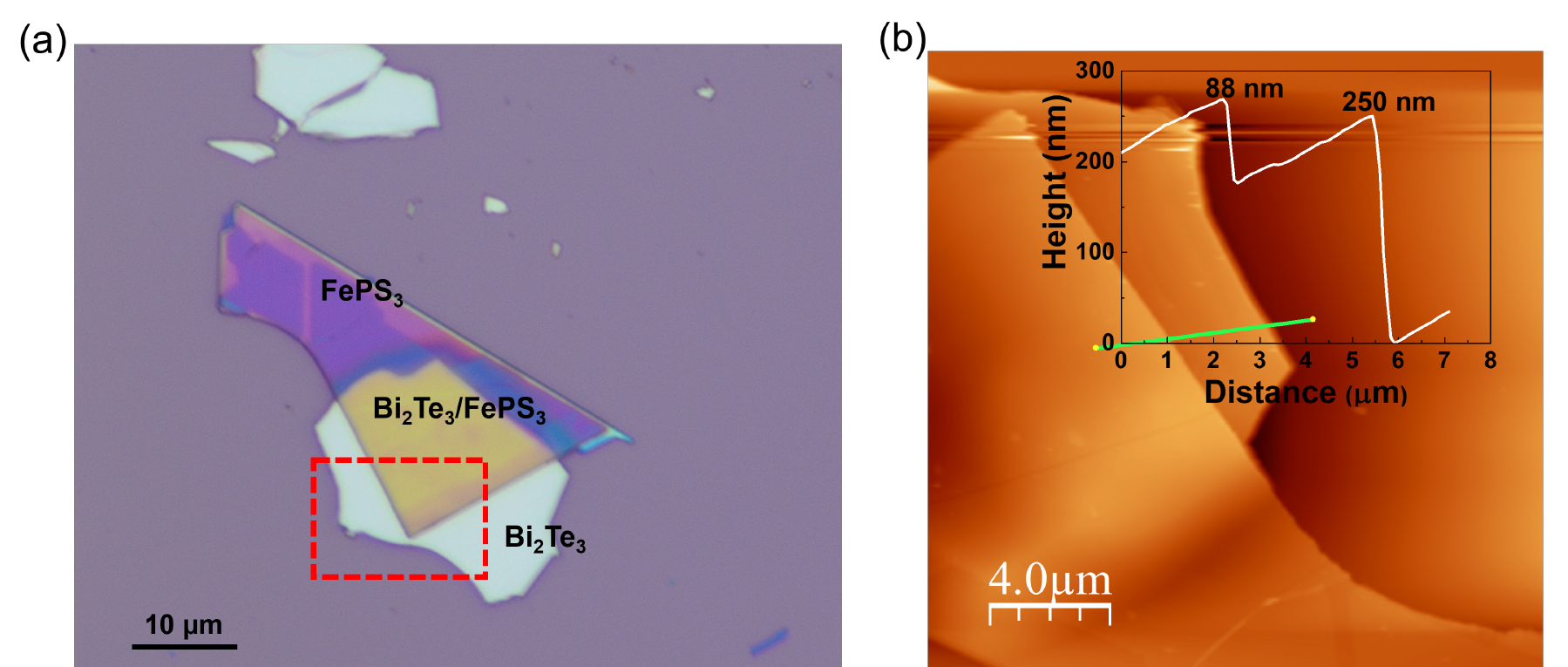}}
\caption{(a) Optical microscopy image of Bi$_{2}$Te$_{3}$/FePS$_{3}$ (HS-2). (b) Atomic force microscopy image of Bi$_{2}$Te$_{3}$/FePS$_{3}$ (HS-2). Inset of (b) shows the height profile of the heterostructure.}
\end{figure}
\begin{figure}[H]
\centerline{\includegraphics[scale=0.5, clip]{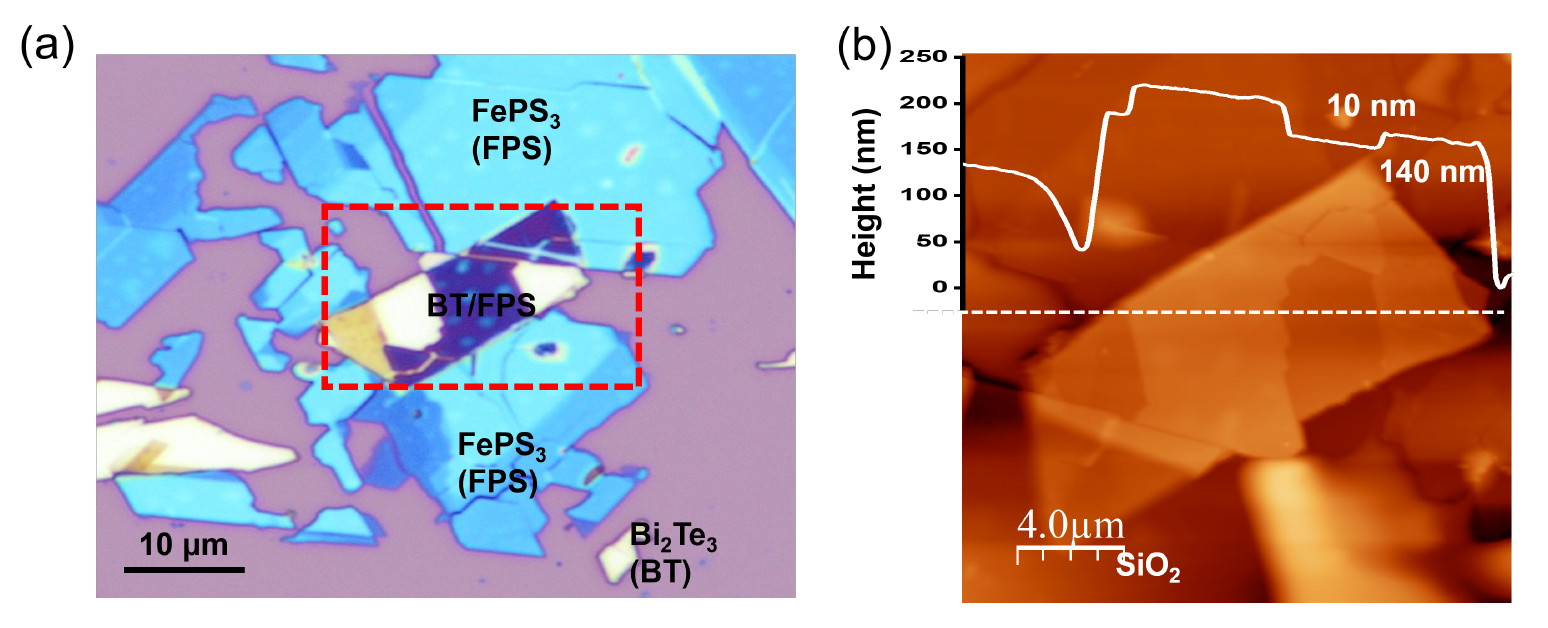}}
\caption{(a) Optical microscopy image of Bi$_{2}$Te$_{3}$/FePS$_{3}$ (HS-3). (b) Atomic force microscopy image of Bi$_{2}$Te$_{3}$/FePS$_{3}$ (HS-3). Inset of (b) shows the height profile of the heterostructure.}
\end{figure}
\begin{figure}[H]
\centerline{\includegraphics[scale=0.5, clip]{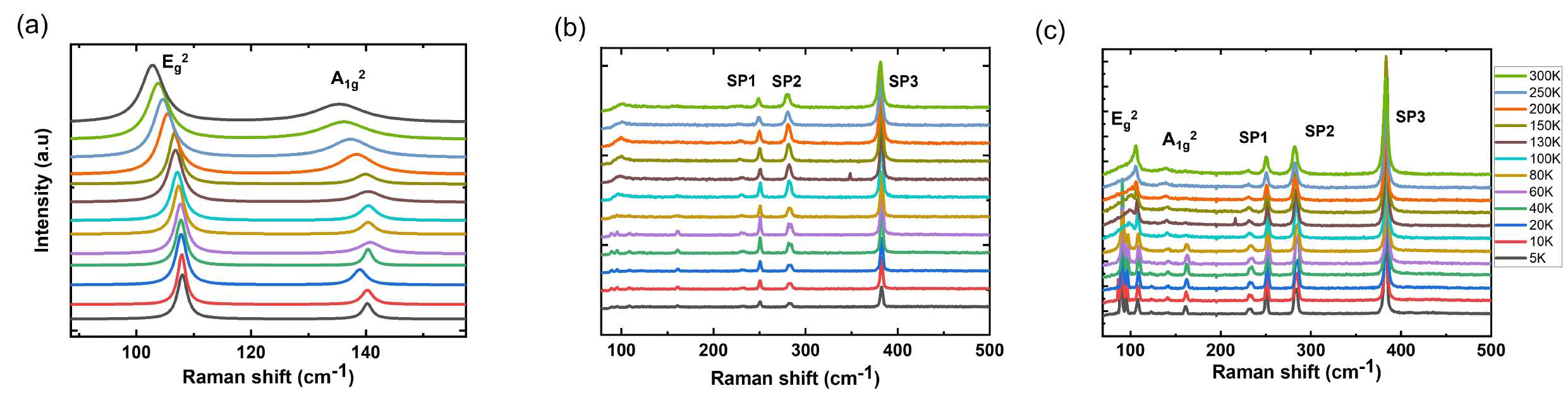}}
\caption{Raman spectra of (a) isolated Bi$_{2}$Te$_{3}$, (b) isolated FePS$_{3}$, (c) Bi$_{2}$Te$_{3}$/FePS$_{3}$ heterojunction of multiple sampling points at different temperature (5K to 300K)}
\end{figure}
\begin{figure}[H]
\centerline{\includegraphics[scale=0.5, clip]{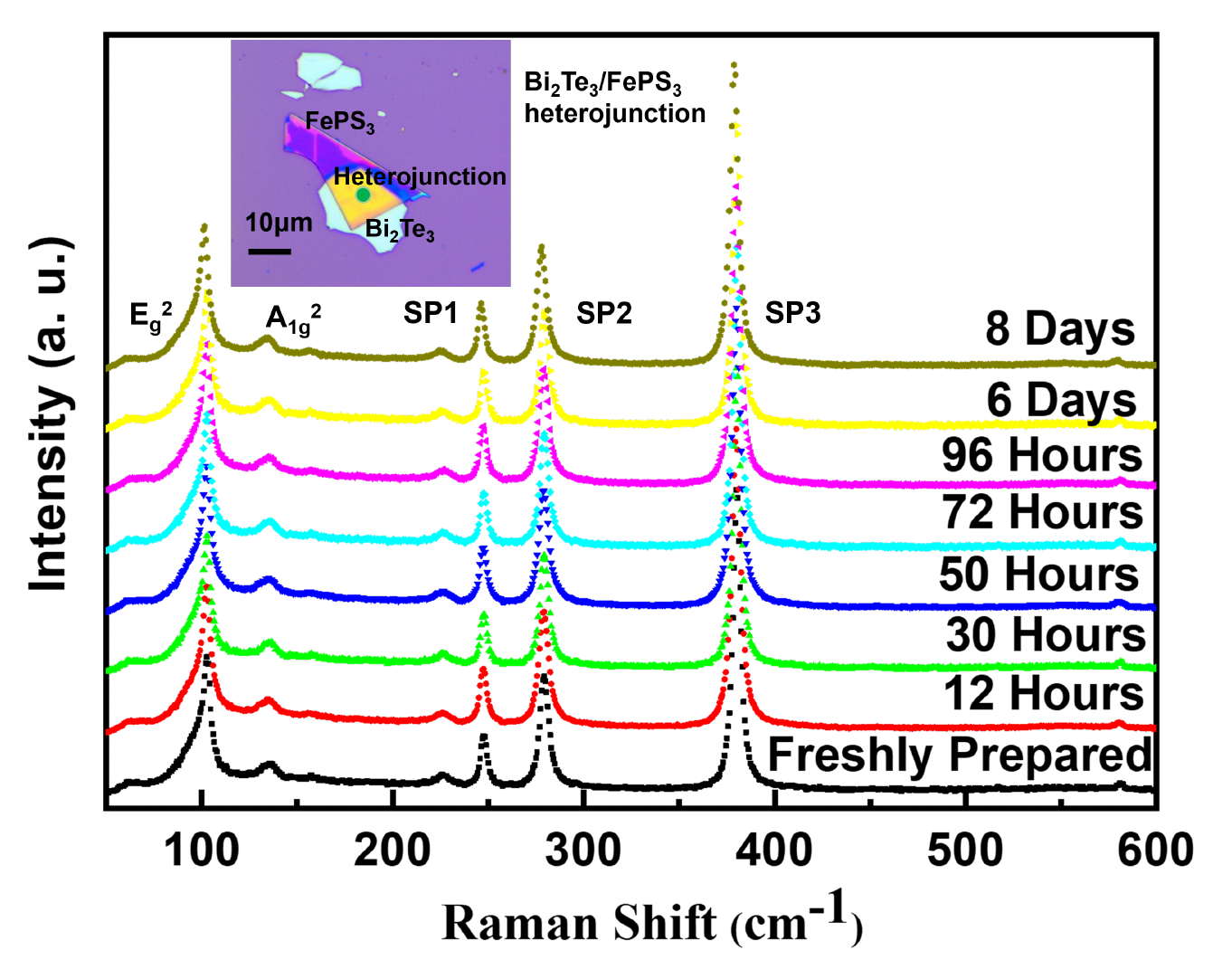}}
\caption{Raman Spectra of Bi$_{2}$Te$_{3}$/FePS$_{3}$ heterojunction were recorded for 8 days to check the air stability of the heterostructure. Inset shows the optical microscopy image of the heterostructure. Green dot in heterojunction signifies the laser spot. No change was observed in peak position and linewidth of Bi$_{2}$Te$_{3}$ and FePS$_{3}$ Raman modes.}
\end{figure}
\begin{figure}[H]
\centerline{\includegraphics[scale=0.5, clip]{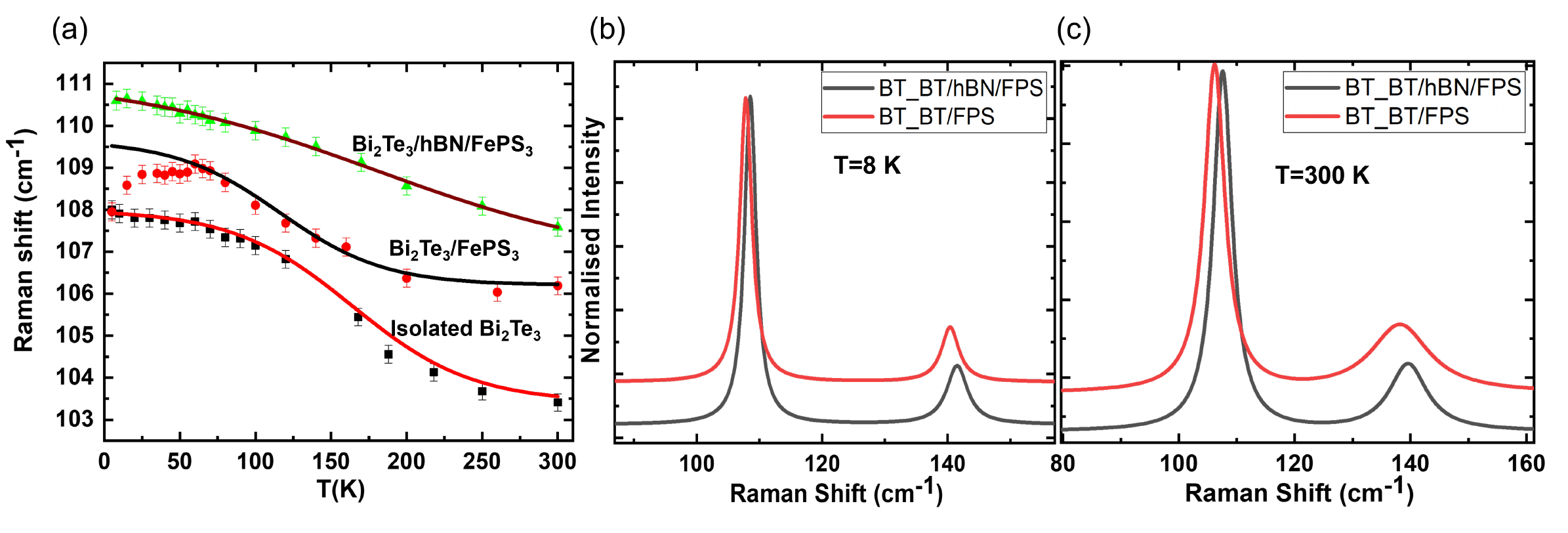}}
\caption{(a) Variation of in-plane Raman modes of Bi$_{2}$Te$_{3}$ are plotted with temperature for isolated and for different heterostructures; (b), (c) Raman shifts of Bi$_{2}$Te$_{3}$ modes are shown for two different temperatures in heterostructures.}
\end{figure}
\begin{figure}[H]
\centerline{\includegraphics[scale=0.5, clip]{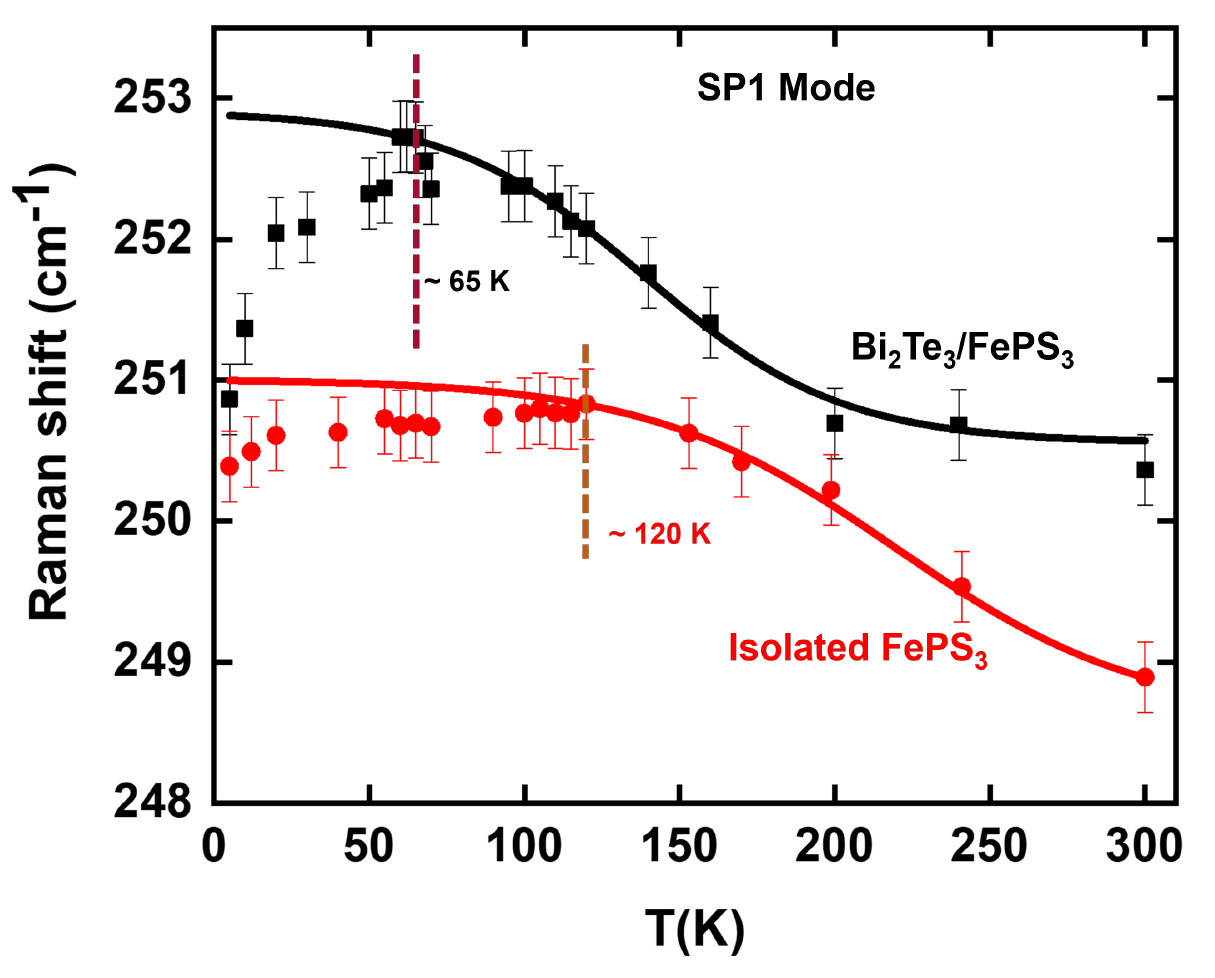}}
\caption{Variation of FePS$_{3}$ spin-phonon coupled mode (SP1) with temperature for two different configurations. In case of isolated FePS$_{3}$, the characteristic N{\'e}el temperature (T$_{N}$) is around 120 K and reduction of T$_{N}$ was observed at/around 65 K due to Bi$_{2}$Te$_{3}$ underneath in heterostructure.}
\end{figure}
\begin{figure}[H]
\centerline{\includegraphics[scale=0.5, clip]{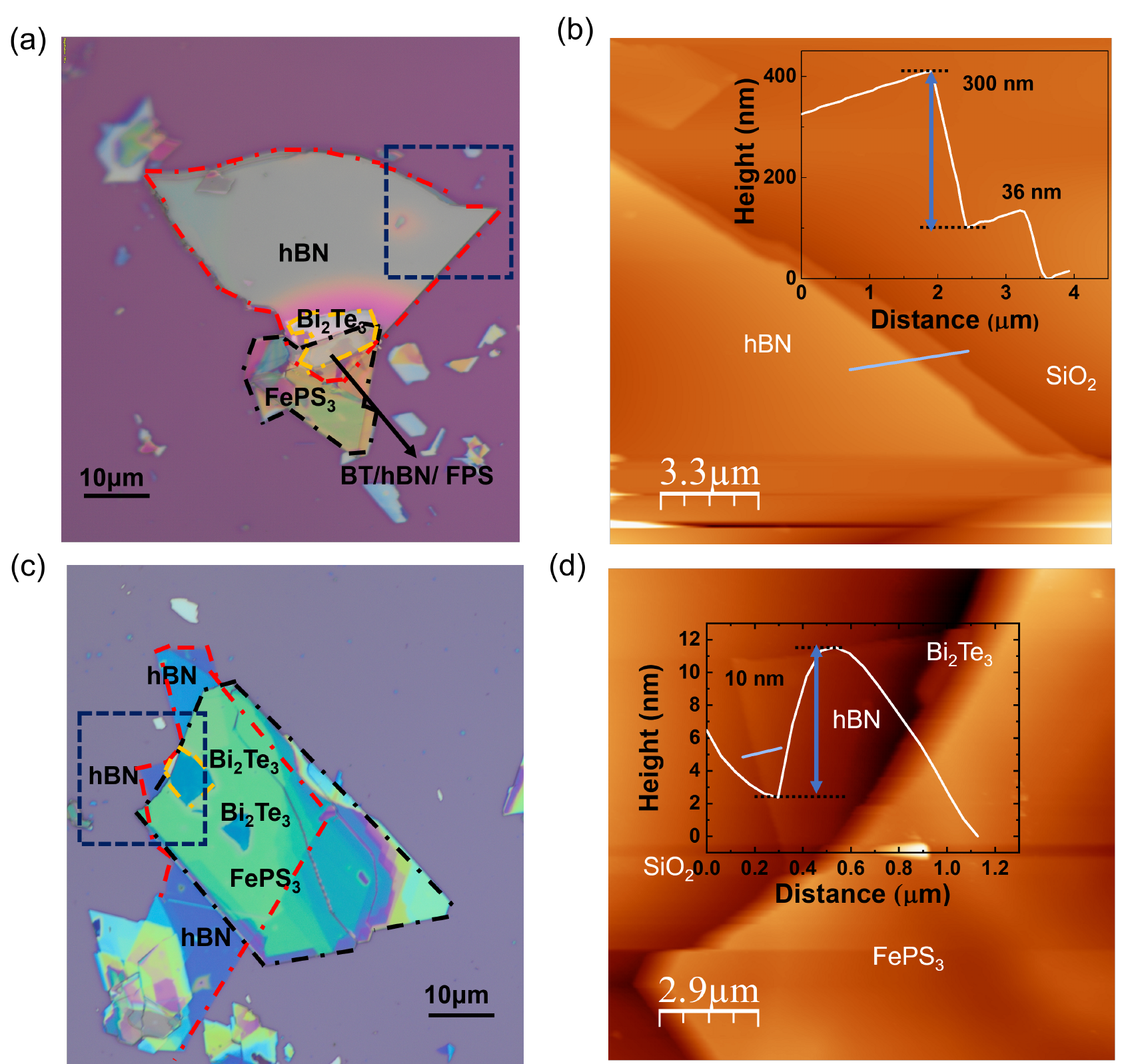}}
\caption{ (a), (c) Optical microscopy images of Bi$_{2}$Te$_{3}$/hBN/FePS$_{3}$ HS with two different thickness of hBN. (b),(d) Atomic force microscopy images of HS. Insets of (b), (d) show the height profile of hBN}
\end{figure}
\begin{figure}[H]
\centerline{\includegraphics[scale=0.5, clip]{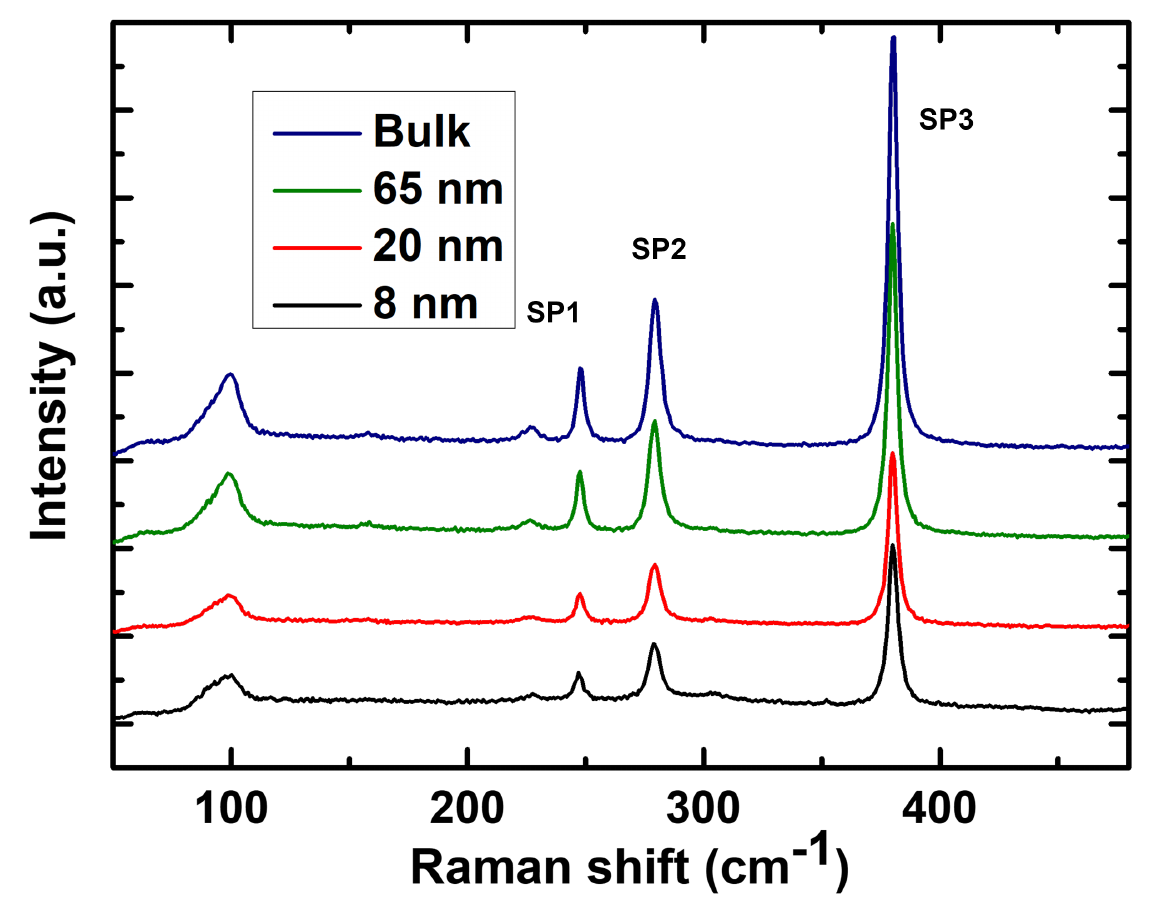}}
\caption{Room temperature Raman spectra of FePS$_{3}$ with varying thickness. No significant Raman shift was observed with thickness indicating layer dependent stability of vibrational modes even in few layer FePS$_{3}$.}
\end{figure}
\begin{figure}[H]
\centerline{\includegraphics[scale=0.6, clip]{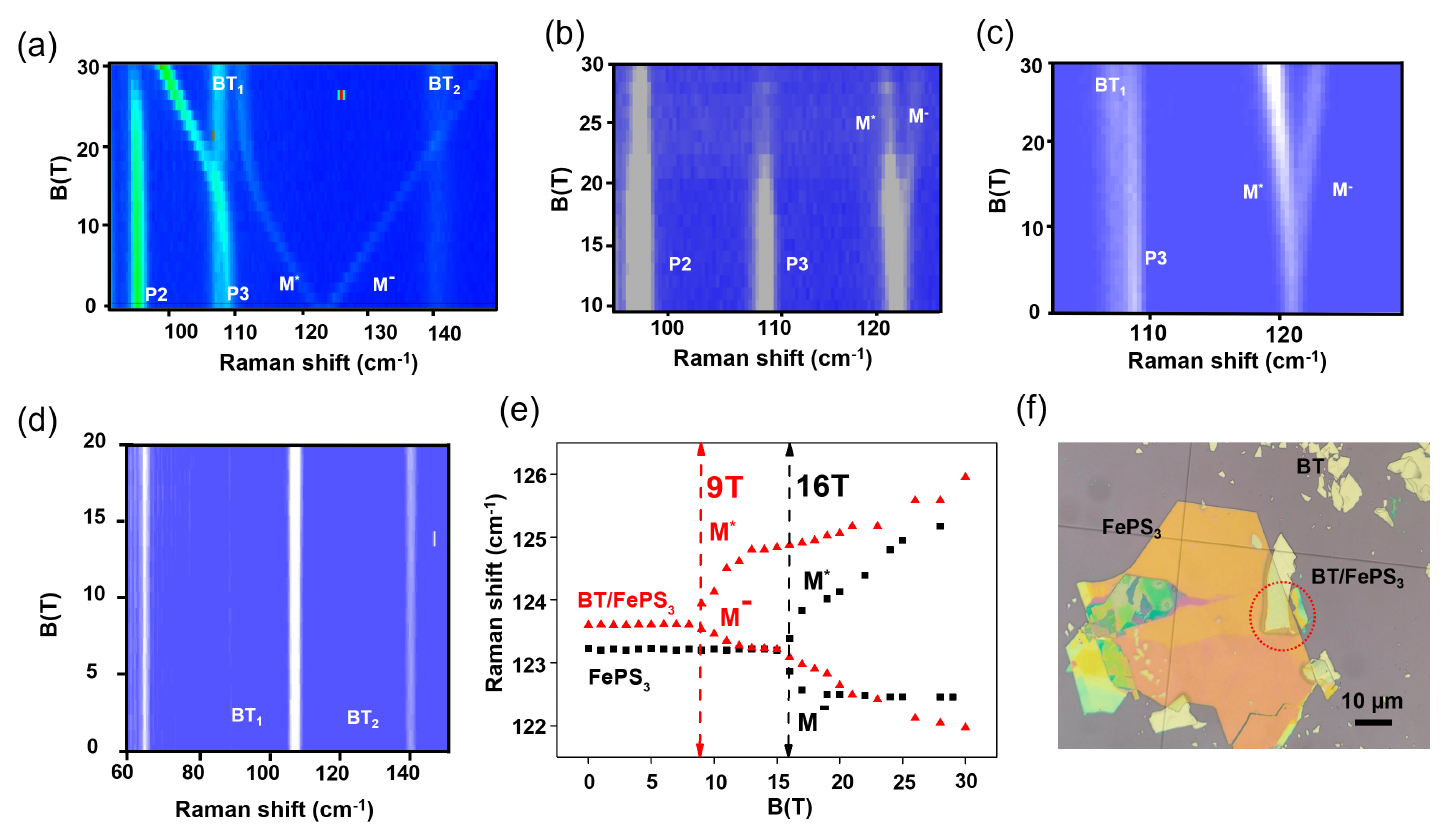}}
\caption{(a) Evolution of the low-temperature (T $\sim$ 4 K) magneto-Raman scattering response of Bi$_{2}$Te$_{3}$/FePS$_{3}$ heterostructure with magnetic field in Faraday geometry, in the spectral region from 80 to 140 cm$^{−1}$. E$_{g}^{2}$ and A$_{1g}^{2}$ phonon modes of Bi$_{2}$Te$_{3}$ are designated as BT$_{1}$ and BT$_{2}$ respectively. Phonon modes (BT$_{1}$, BT$_{2}$) of both Bi$_{2}$Te$_{3}$ and FePS$_{3}$ (P2, P3, Magnon (M)) are visible in the colour map. The magnon mode  at 120 cm$^{-1}$ splits into two components M$^{*}$ and M$^{-}$ at/around 1 T. (b) Raman modes of FePS$_{3}$ with magnetic field in Voigt configuration at T$\sim$ 4K. The magnon mode splits at/around 16 T. (c) Raman modes of Bi$_{2}$Te$_{3}$/FePS$_{3}$ heterostructure with magnetic field in Voigt configuration at T $\sim$ 4 K. The magnon mode in FePS$_{3}$ splits at lower field ($\sim$ 9 T) compared to isolated FePS$_{3}$. (d) No magnetic field dependence was observed in the phonon modes of Bi$_{2}$Te$_{3}$. (e) Magnon frequency was plotted with magnetic field for both isolated FePS$_{3}$ and Bi$_{2}$Te$_{3}$/FePS$_{3}$ heterostructures in Voigt geometry. The magnon splitting occurs much lower field ($\sim$ 9 T) in heterostructure compared to splitting field in isolated FePS$_{3}$ ($\sim$ 16 T) (f)  Optical microscopy image of Bi$_{2}$Te$_{3}$/FePS$_{3}$ heterostructure. High magnetic field Raman spectroscopy was performed in this heterostructure. The area marked with dotted red circle indicates the heterostructure region.}
\end{figure}
\begin{figure}[H]
\centerline{\includegraphics[scale=0.45, clip]{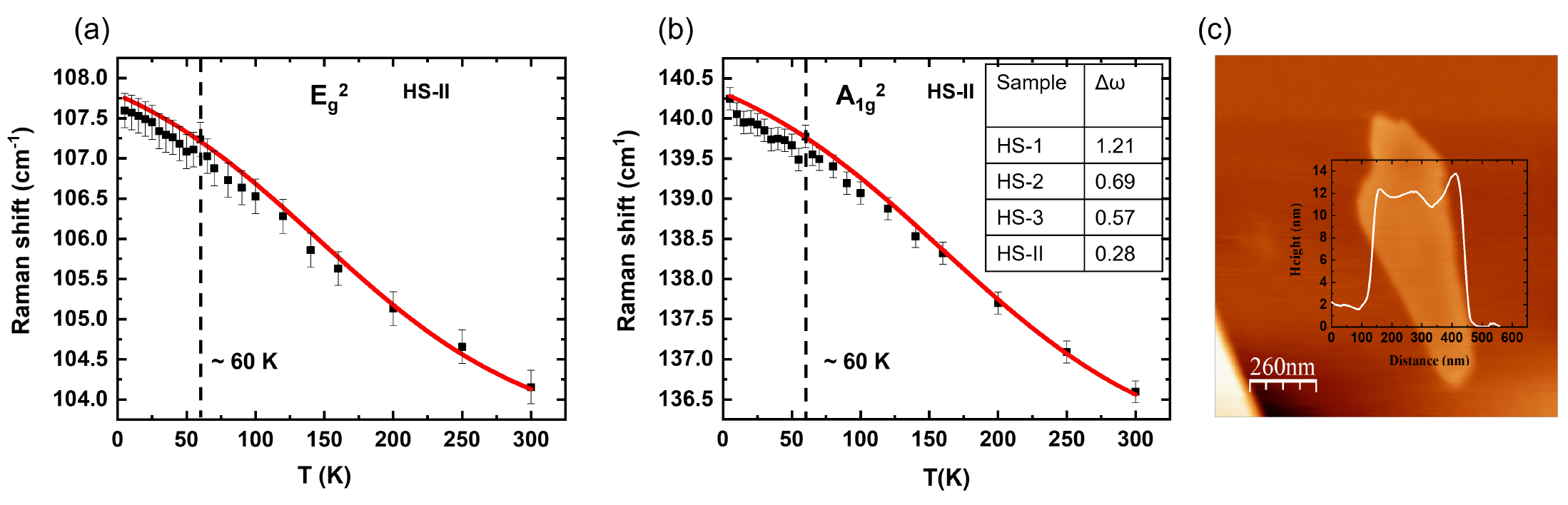}}
\caption{Temperature dependent Raman modes of Bi$_{2}$Te$_{3}$ in HS-II. (a) In-plane (E$_{g}^{2}$) phonon mode; (b) out-of-plane (A$_{1g}^{2}$) phonon mode. Insets of figure (b) shows the reduction of $\Delta\omega$ with heterostructure thickness. HS (1-3) was already mentioned in the manuscript. HS-II is freshly prepared heterostructure. (c) Atomic force microscopy (AFM) image of thinner Bi$_{2}$Te$_{3}$ used in fabricating new heterostructures (HS-I, HS-II). The thickness of the flake is $\sim$ 12 nm confirmed by the height profile.}
\end{figure}
\centering
\begin{table}[h]
\begin{tabular}{|c|c|c|c|}
\cline{1-3}
\multicolumn{3}{c}{Bi$_{2}$Te$_{3}$}\\
\cline{1-3}
Raman Modes &  $\omega (cm^{-1})$(Expt) & $\omega (cm^{-1})$(Theory) \\ \cline{1-3}
E$_{g}$ & - & 36 \\ \cline{1-3}
A$_{1g}$ & - & 54 \\ \cline{1-3}
E$_{g}$ & \textbf{106} & \textbf{104} \\ \cline{1-3}
A$_{1g}$ & \textbf{138} & \textbf{138} \\ \cline{1-3}
\end{tabular}
\centering \caption{Irreducible representations of the Raman-active modes of bulk Bi$_{2}$Te$_{3}$. Only the high frequency E$_{g}$ and A$_{1g}$ modes are observed in experiments (second column).}
\end{table}

\begin{table}[h]
\centering
\begin{tabular}{|c|c|c|c|c|}
\hline
\multicolumn{3}{c}{FePS$_{3}$}\\
\hline
Raman Modes &  $\omega (cm^{-1})$(zigzag) & $\omega (cm^{-1})$(FM) & $\Delta^{rel}_{\lambda}(\%)$ \\ \hline
B$_{g}$ & 97 & 95 & -2.46 \\ \hline
A$_{g}$ & \textbf{115} & \textbf{102} & -11.56 \\ \hline
A$_{g}$ & \textbf{141} & \textbf{137}& -2.97 \\ \hline
B$_{g}$ & 140 & 146 & 3.71 \\ \hline
B$_{g}$ & 203 & 205 & 0.74 \\ \hline
A$_{g}$ & 208 & 211 & 1.73 \\ \hline
B$_{g}$ & 214 & 214 & 0.09 \\ \hline
A$_{g}$ & 225 & 230 & 2.31 \\ \hline
B$_{g}$ & 256 & 259 & 0.90 \\ \hline
A$_{g}$ & 268 & 268 & 0.04 \\ \hline
A$_{g}$ & 356 & 359  & 0.84 \\ \hline
A$_{g}$ & 536 & 538 & 0.32 \\ \hline
B$_{g}$ & 529 & 529 & 0.04 \\ \hline
A$_{g}$ & 546 & 547 & 0.18 \\ \hline
\end{tabular}
\caption{Irreducible representations of the Raman-active modes of FePS$_{3}$. Phonon frequencies are calculated for zig-zag AFM (z-AFM) and FM spin configurations. Relative angular frequency shift ($\Delta^{rel}_{\lambda}= \frac{\omega_{FM}-\omega_{z-AFM}}{\omega_{z-AFM}}\times100\%$) due to change in the magnetic ordering of the Raman-active modes is given in the last column.
 }
\end{table}

\begin{figure}[H]
\centerline{\includegraphics[scale=0.9, clip]{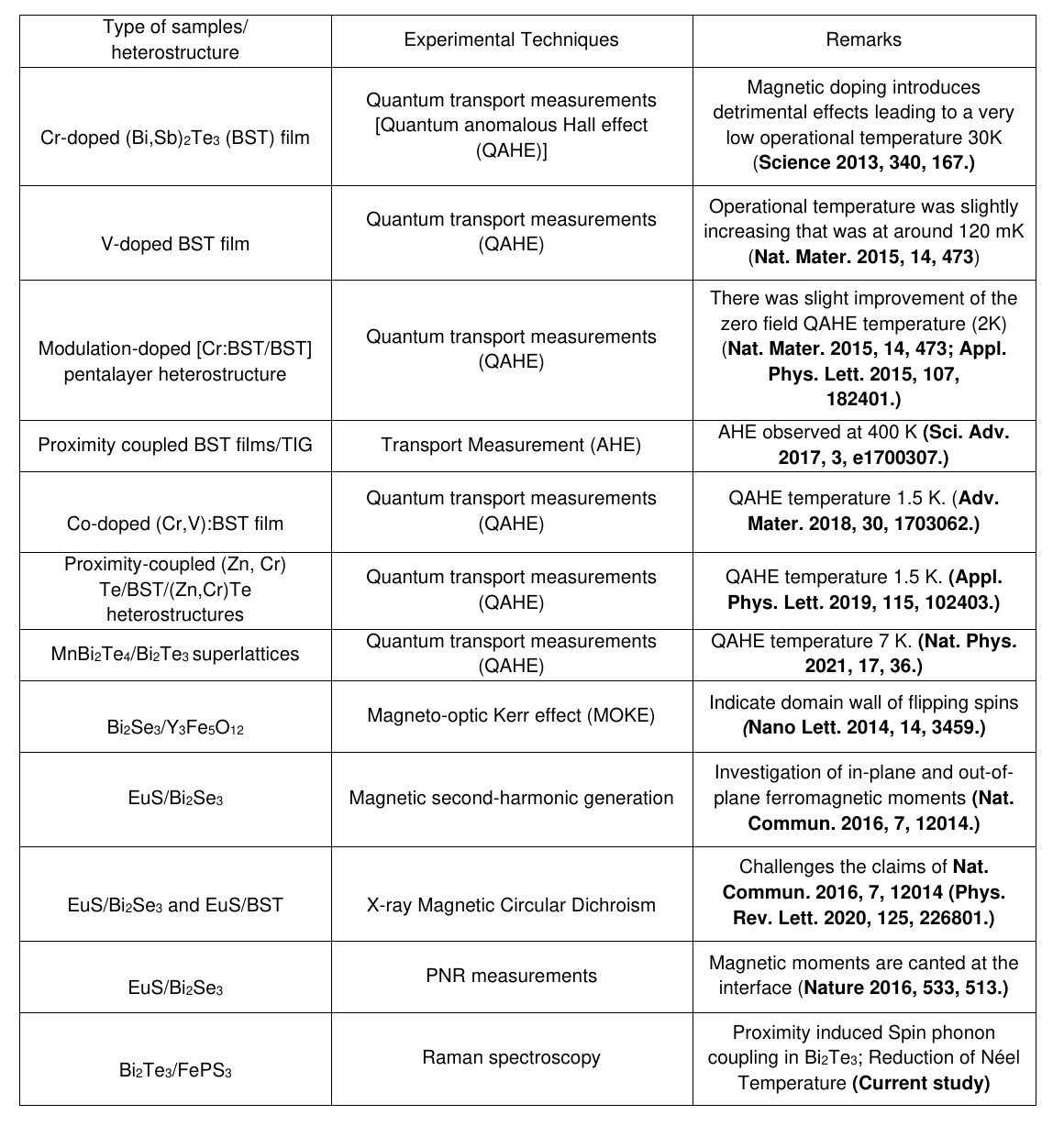}}
  TABLE III.  Performance comparisons of different Topological Insulator heterostructures using various experimental techniques
\end{figure}
\begin{figure}[H]
\centerline{\includegraphics[scale=0.85, clip]{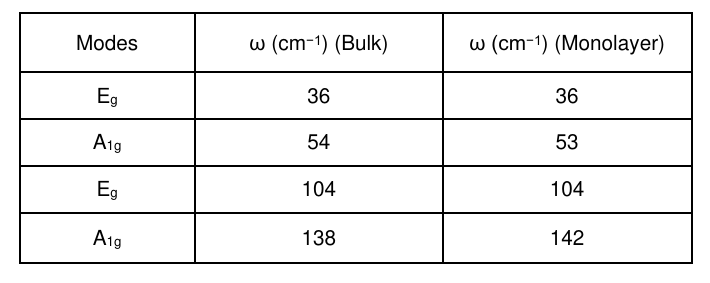}}
  \centerline{TABLE IV. Irreducible representations of the Raman-active modes of bulk and single layer Bi$_{2}$Te$_{3}$.} 
\end{figure}
\begin{figure}[H]
\centerline{\includegraphics[scale=0.85, clip]{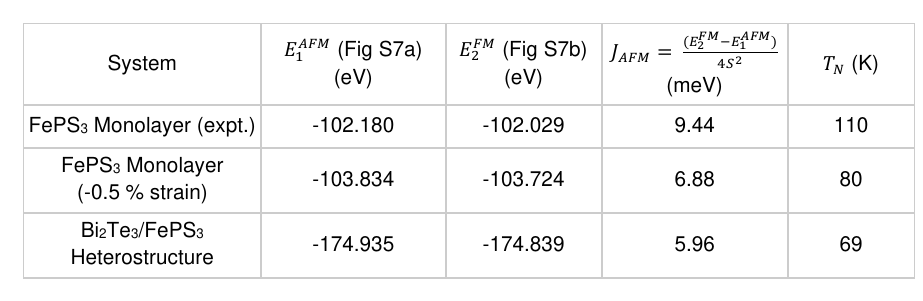}}
  TABLE V.  Calculation of AFM exchange in FePS$_{3}$ monolayer 
  with and without strain, and within the heterostructure setup involving Bi$_{2}$Te$_{3}$.
\end{figure}
\begin{figure}[H]
\centerline{\includegraphics[scale=0.85, clip]{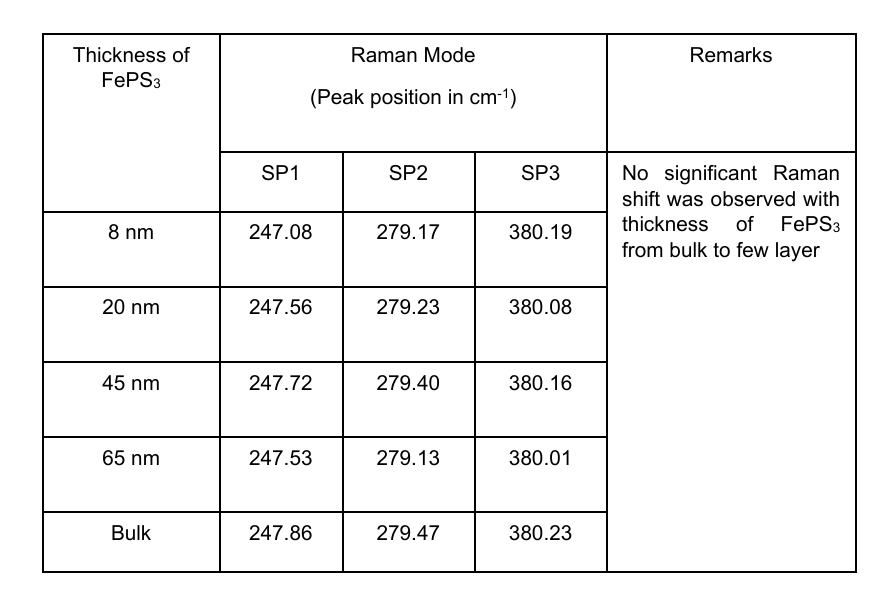}}
  TABLE VI.  The frequencies of FePS$_{3}$ spin-phonon (SP) coupled modes are presented with varying thickness (Figure S23). No significant Raman shift was observed with the thickness of FePS$_{3}$.
\end{figure}
\begin{figure}[H]
\centerline{\includegraphics[scale=0.85, clip]{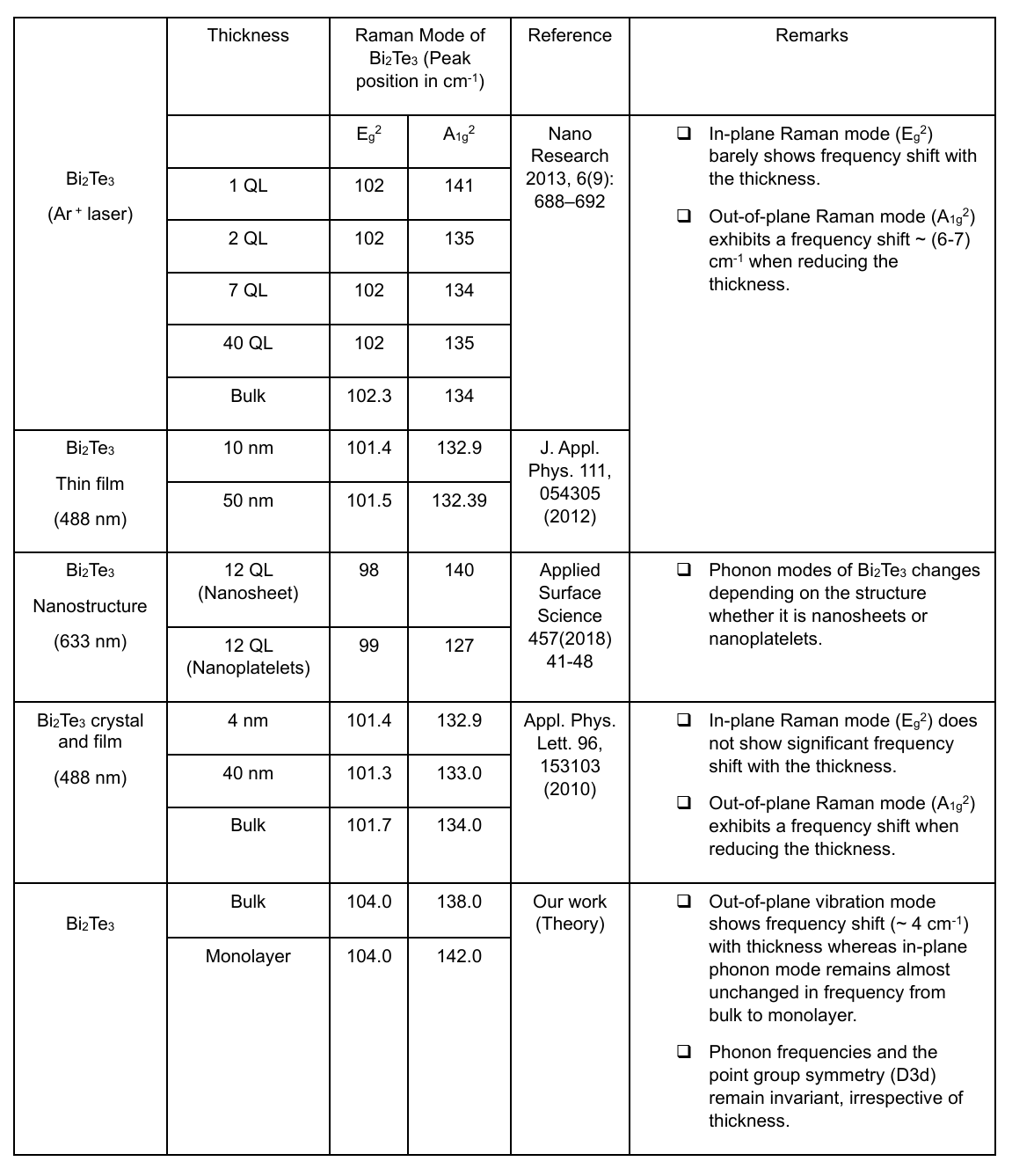}}
  TABLE VII. Thickness dependent Raman shift of Bi$_{2}$Te$_{3}$ with different structural form and wavelength excitation. No significant Raman shift was observed in E$_{g}^{2}$ mode and a prominent shift ($\sim $ 6 cm$^{-1}$) was observed in A$_{1g}^{2}$ mode with thickness.
\end{figure}
\begin{figure}[H]
\centerline{\includegraphics[scale=0.85, clip]{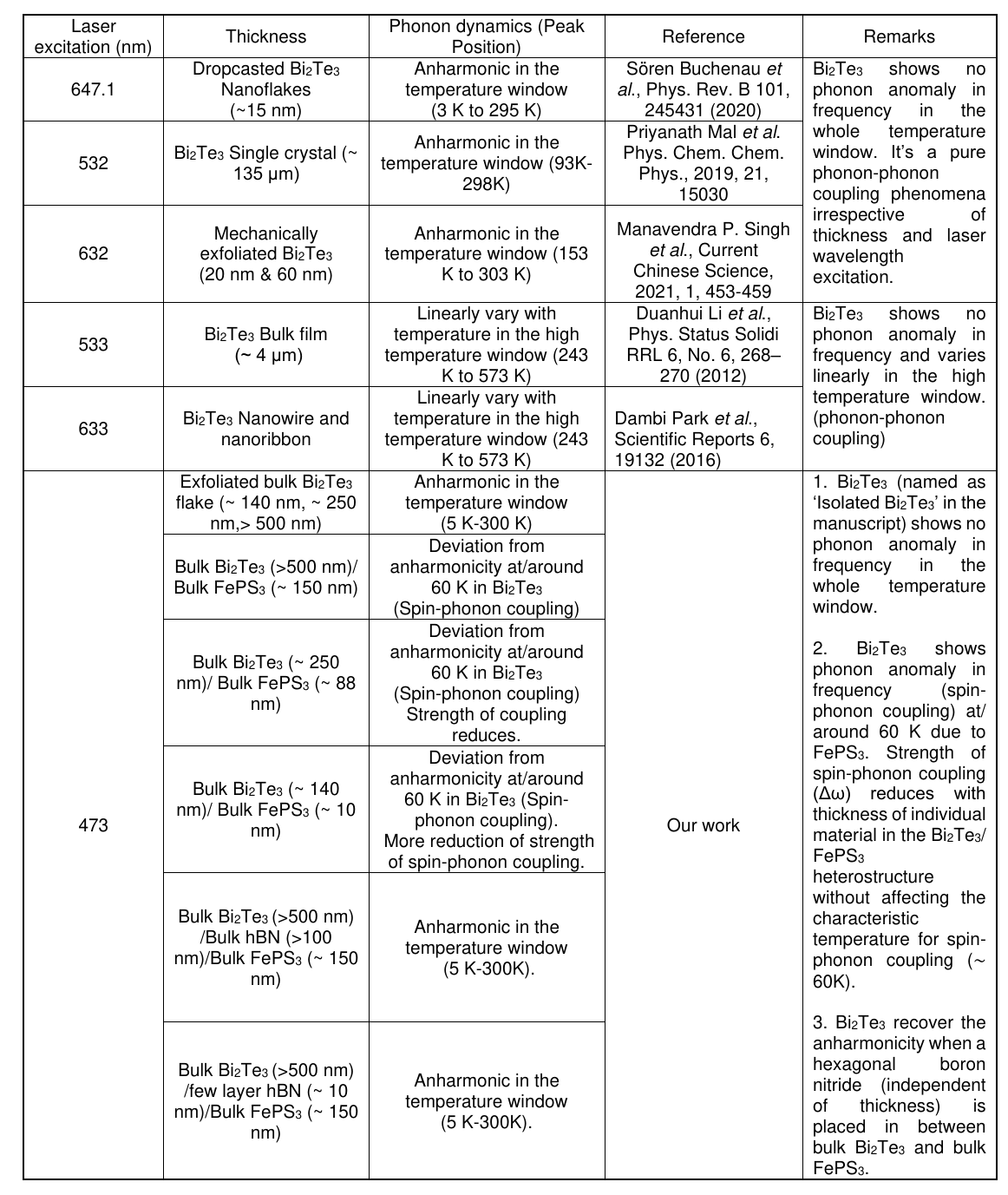}}
  TABLE VIII. Phonon dynamics of  Bi$_{2}$Te$_{3}$ with temperature for a fixed thickness of Bi$_{2}$Te$_{3}$. Bi$_{2}$Te$_{3}$ shows no phonon anomaly in frequency with temperature irrespective of thickness, wavelength excitation and different form of Bi$_{2}$Te$_{3}$. It is pure phonon-phonon coupling.      
\end{figure}
\begin{figure}[H]
\centerline{\includegraphics[scale=0.85, clip]{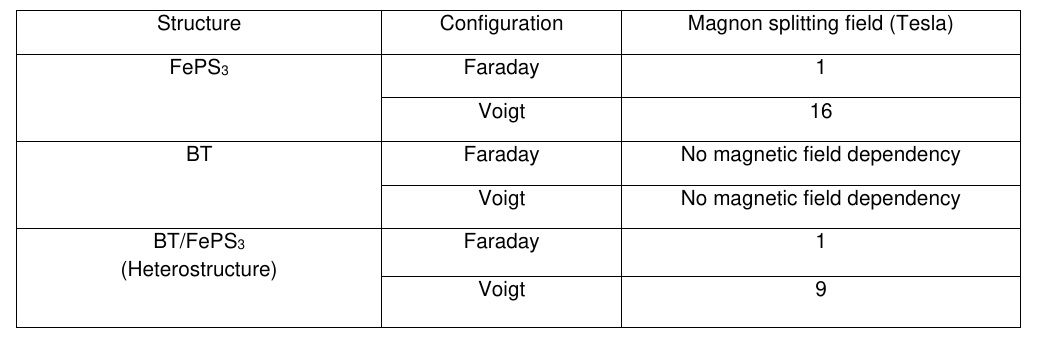}}
  TABLE IX. Magnon splitting field in FePS$_{3}$, Bi$_{2}$Te$_{3}$ and Bi$_{2}$Te$_{3}$/FePS$_{3}$ heterostructures in different magnetic field configurations.      
\end{figure}